\documentclass[reprint,superscriptaddress,aps,prl,twocolumn]{revtex4-2}
\usepackage[T1]{fontenc}
\usepackage[latin1]{inputenc}
\usepackage{lmodern}
\expandafter\let\csname equation*\endcsname\relax
\expandafter\let\csname endequation*\endcsname\relax
\usepackage{amsmath}
\usepackage{amssymb}
\usepackage{graphicx}
\usepackage{xcolor}
\usepackage{natbib}
\usepackage{bm}
\usepackage{bbold}
\usepackage[mathscr]{euscript}
\usepackage[cal=boondoxo]{mathalfa}
\usepackage{comment}
\renewcommand\footnotemark{}

\usepackage{braket}
\usepackage[normalem]{ulem}

\def\ii{{\rm i}}

\def\db{\boldsymbol{\wp}} 
  
\def\rb{{\bf r}}

\def\hge{\hat{\sigma}_{ge}}  
\def\heg{\hat{\sigma}_{eg}}

\def\bra#1{\mathinner{\langle{#1}|}}
\def\ket#1{\mathinner{|{#1}\rangle}}
\def\braket#1{\mathinner{\langle{#1}\rangle}}
\def\jop{\hat{\mathcal{O}}}
\def\dop{\hat{\mathcal{D}}}
\def\nexc{\hat{n}_{\mathrm{exc.}}}
\def\bnexc{\braket{\hat{n}_{\mathrm{exc.}}(t)}}
\newcommand\varpm{\mathbin{\vcenter{\hbox{%
  \oalign{\hfil$\scriptstyle+$\hfil\cr
          \noalign{\kern-.3ex}
          $\scriptscriptstyle({-})$\cr}%
}}}}

\begin{document}
\title{Many-body Signatures of Collective Decay in Atomic Chains}
\author{Stuart J. Masson}
\email{s.j.masson@columbia.edu}
\affiliation{Department of Physics, Columbia University, New York, New York 10027, USA}
\author{Igor Ferrier-Barbut}\affiliation{Universit\'{e} Paris-Saclay, Institut d'Optique Graduate School, CNRS, Laboratoire Charles Fabry, 91127, Paris, France}
\author{Luis A. Orozco}
\affiliation{Joint Quantum Institute, Department of Physics and NIST, University of Maryland, College Park, Maryland 20742, USA}
\author{Antoine Browaeys}
\affiliation{Universit\'{e} Paris-Saclay, Institut d'Optique Graduate School, CNRS, Laboratoire Charles Fabry, 91127, Paris, France}
\author{Ana Asenjo-Garcia}
\email{ana.asenjo@columbia.edu}
\affiliation{Department of Physics, Columbia University, New York, New York 10027, USA}

\date{\today}

\begin{abstract}
    Fully inverted atoms placed at exactly the same location synchronize as they deexcite, and light is emitted in a burst (known as ``Dicke's superradiance''). We investigate the role of finite interatomic separation on correlated decay in mesoscopic chains, and provide an understanding in terms of collective jump operators. We show that the superradiant burst survives at small distances, despite Hamiltonian dipole-dipole interactions. However, for larger separations, competition between different jump operators leads to dephasing, suppressing superradiance. Collective effects are still significant for arrays with lattice constants of the order of a wavelength, and lead to a photon emission rate that decays nonexponentially in time. We calculate the two-photon correlation function and demonstrate that emission is correlated and directional, as well as sensitive to small changes in the interatomic distance. These features can be measured in current experimental setups, and are robust to realistic imperfections.
\end{abstract}

\maketitle

Collective effects in the interaction between light and matter have attracted interest since the seminal work of Dicke in the 1950s, who studied the problem of photon emission by many atoms at identical locations~\cite{Dicke54}. In this purely dissipative scenario, the atomic dipoles become phase locked during their decay and emit collectively. This leads to an initial increase in the photon emission rate -- the famous ``superradiant burst'' or ``superfluorescence'' -- rather than the typical exponential decay for independent atoms.

\begin{figure}
    \includegraphics[width=0.48\textwidth]{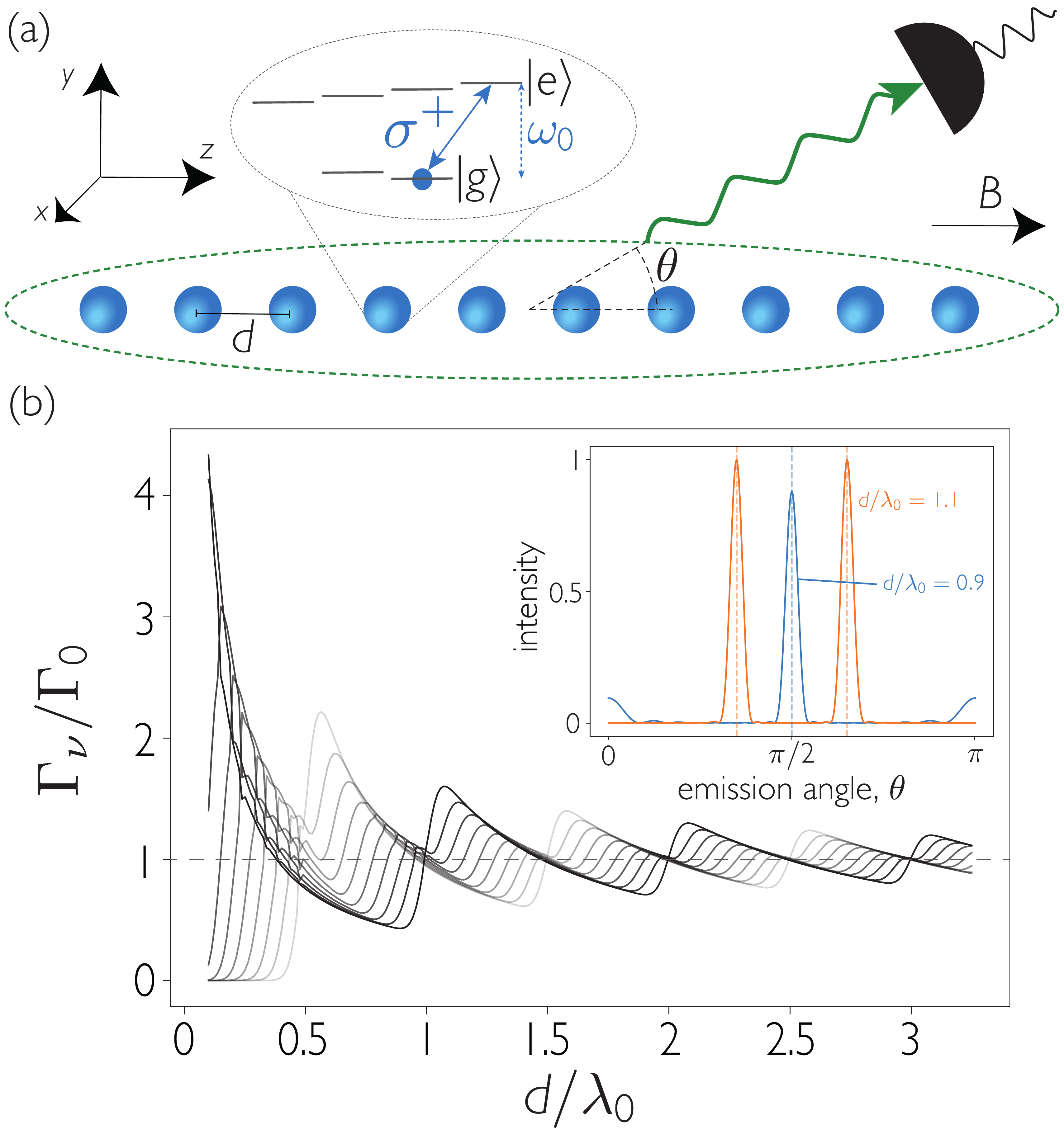}
    \caption{\textbf{A chain of excited atoms decays collectively, emitting correlated photons.} \textbf{(a)}~Schematic of the considered setup. Atoms have resonance frequency $\omega_0$, and are separated by a constant distance $d$. The relevant transition is selected via a small magnetic field. \textbf{(b)}~Decay rates of the jump operators $\set{\jop_\nu}$ for $N=10$ atoms. Each operator approximates a spin wave with wave vector $k_\nu$, with the darkest (lightest) lines corresponding to spin waves with minimum (maximum) wave vector. Inset: Angular emission pattern (in arbitrary units) following action of the most subradiant operator on a fully inverted array, measured by detectors of width $\Delta\theta = 0.01\pi$. Dashed lines are analytically obtained angles of peak emission, $\theta_{\mathrm{max.}} = \mathrm{arccos}(\pm k_\nu / k_0)$ with $k_\nu=0(\pi/d)$ for $d=0.9 (1.1)\lambda_0$.}
    \label{Fig1}
\end{figure}

Dicke's scenario ignores coherent dipole-dipole interactions between atoms, which are relevant for finite interatomic distances and have been predicted to wash out superradiant decay~\cite{Gross82,BenedictBook}. However, signatures of collective behavior persist even in systems of size much larger than the resonance wavelength. For example, theoretical studies of ordered arrays of emitters have shown the existence of extremely subradiant (i.e., dark) few-excitation states~\cite{Zoubi10,Bettles16PRL,Sutherland16,Bettles16PRA,Facchinetti16,Asenjo17PRX,Henriet19,Masson20arxiv,Williamson20PRL}, as well as directional collective emission~\cite{Carmichael00,Clemens03,Clemens04,Scully06,Zoubi10,Bhatti15,Shahmoon17,Gulfam18,Liberal19,Ballantine20}.

Recent experimental realizations of ordered atomic arrays, both in optical lattices~\cite{Bakr10,Sherson10,Greif16,Kumar18,Rui20} and tweezer arrays~\cite{Lester15,Kim16,Endres16,Barredo16,Barredo18,Browaeys20,Glicenstein20}, open the door for investigation of these predictions. These platforms have already allowed for the demonstration of a two-dimensional atomic mirror~\cite{Rui20} and the measurement of collective frequency shifts in a one-dimensional (1D) atomic array~\cite{Glicenstein20}. Current experimental capabilities offer the possibility of measuring statistics of the emitted photons. This raises the question of whether collective decay imprints correlations on the emitted photons. This would allow for the potentially tunable generation of nonclassical states of light (and maybe of a superradiant laser~\cite{Meiser09,Bohnet12}), critical for quantum technologies.  Conversely, connecting the correlations in the light back to the atomic quantum state would offer a unique light-based probe to characterize these dissipative many-body systems. 

Here, we make an important step in this direction by investigating collective decay and superradiance in mesoscopic 1D ordered arrays. We find that a superradiant burst survives for short interatomic distances ($d\lesssim\lambda_0/4$). We use a quantum jump approach to connect the sequence of collective jumps to the statistics of the light emitted. We show that strong spatial correlations between emitted photons are imparted by the collective decay and that they persist even for $d\sim\lambda_0$.
 
 \color{black}

We consider $N$ atoms of resonance frequency $\omega_0$, arranged in an ordered chain along the $z$-axis with lattice constant $d$, as shown in Fig.~\ref{Fig1}(a). Interactions between atoms are obtained by tracing out the electromagnetic field under a Born-Markov approximation~\cite{Gruner96,Dung02}. The atomic density matrix, $\rho$, evolves under the master equation
\begin{equation}\label{masterequation}
\dot{\rho} = - \frac{\ii}{\hbar} [\mathcal{H},\rho] + \sum\limits_{i,j=1}^N \frac{\Gamma^{ij}}{2} \left( 2\hat{\sigma}_{ge}^j \rho \hat{\sigma}_{eg}^i - \rho \hat{\sigma}_{eg}^i\hat{\sigma}_{ge}^j - \hat{\sigma}_{eg}^i \hat{\sigma}_{ge}^j \rho \right),
\end{equation}
where the Hamiltonian is
\begin{equation}\label{hamiltonian}
\mathcal{H} = \hbar\sum_{i=1}^N\omega_0\hat{\sigma}_{ee}^i + \hbar\sum_{i,j=1}^N J^{ij}\hat{\sigma}_{eg}^i\hat{\sigma}_{ge}^j.
\end{equation}
Here, $\hat{\sigma}_{ge}^i = \ket{g_i} \bra{e_i}$ is the atomic coherence operator, with $\ket{e_i}$ and $\ket{g_i}$ the excited and ground states of the cycling transition of the $i$th atom at position $\rb_{i} = \set{x_i,y_i,z_i}$. The coherent and dissipative interactions between atoms $i$ and $j$ are~\cite{Stephen64,Lehmberg70}
\begin{equation}\label{shiftrate}
J^{ij} - \ii\frac{\Gamma^{ij}}{2} =-\frac{\mu_0\omega_0^2}{\hbar}\,\db^*\cdot \mathbf{G}_0(\rb_i,\rb_j,\omega_0)\cdot\db,
\end{equation}
where $\db = (|\db|/\sqrt{2})(\hat{x} + \ii \hat{y})$ is the dipole matrix element of the circularly polarized transition $\sigma^+$, and $\mathbf{G}_0(\rb_i,\rb_j,\omega_0)$ is the propagator of the electromagnetic field between positions $\rb_{i}$ and $\rb_{j}$~\cite{Asenjo17PRX}. The scattered field along the axis of the chain is $\sigma^+$ polarized  and the atoms behave as two-level systems even in the presence of complex hyperfine structure~\cite{Hebenstreit17,PineiroOrioli19,Asenjo19}.

An ensemble of $N$ atoms decays collectively, via a set of $N$ jump operators, $\set{\jop_\nu}$, with associated decay rates $\set{\Gamma_\nu}$. These operators are eigenstates of the dissipative interaction matrix $\bf{\Gamma}$ with elements $\Gamma^{ij}$~\cite{Carmichael00,Clemens03,Clemens04}. The master equation can be written in terms of these operators as
\begin{equation}
\dot{\rho} = - \frac{\ii}{\hbar} [\mathcal{H},\rho] + \sum\limits_{\nu=1}^N \frac{\Gamma_\nu}{2} \left( 2\jop_\nu \rho \jop_\nu^\dagger - \rho \jop_\nu^\dagger\jop_\nu - \jop_\nu^\dagger\jop_\nu \rho \right). \label{mejumps}
\end{equation}
The operators' decay rates can be superradiant, i.e., $\Gamma_\nu > \Gamma_0$, or subradiant, i.e., $\Gamma_\nu < \Gamma_0$, with $\Gamma_0\equiv\Gamma^{ii}$ the single-atom spontaneous emission rate.

Jumps happen stochastically and operators act on all atoms with a set of amplitudes and phases sensitive to $d$ and the atomic quantization axis. The states $\set{\jop_\nu^\dagger \ket{g}^{\otimes N}}$ $\left(\set{\jop_\nu \ket{e}^{\otimes N}}\right)$ form an orthonormal basis for the single-excitation (``single-hole'') system. The jump operators can be classified according to the symmetries of the Lindblad operator. For an infinite 1D array, these are discrete translations along $\hat{z}$. Thus, the operators correspond to Bloch waves with a wave vector along the chain direction $k_\nu$, i.e., $\jop_\nu=(1/\sqrt{N})\sum\nolimits_{i=1}^{N} e^{-\ii k_\nu z_i}\hat{\sigma}_{ge}^i$. In 1D geometries, the collective decay rates $\Gamma_\nu$ change with $d$ featuring sharp oscillations at  $d\simeq n\lambda_0$/2~\cite{Nienhuis87,DeVoe96,Bettles16PRA}, as shown in Fig.~\ref{Fig1}(b). For $d = n\lambda_0/2+\epsilon$ ($d = n\lambda_0/2 -\epsilon$),  with $\epsilon\rightarrow0^+$, there are a small number of superradiant (subradiant) operators, with the majority of rates weakly subradiant (superradiant). These oscillations arise from 1D lattice sums and can be understood by analogy with the decay of a dipole in a cavity~\cite{Bettles16PRA}.

\begin{figure*}[t!]
\centerline{\includegraphics[width=\linewidth]{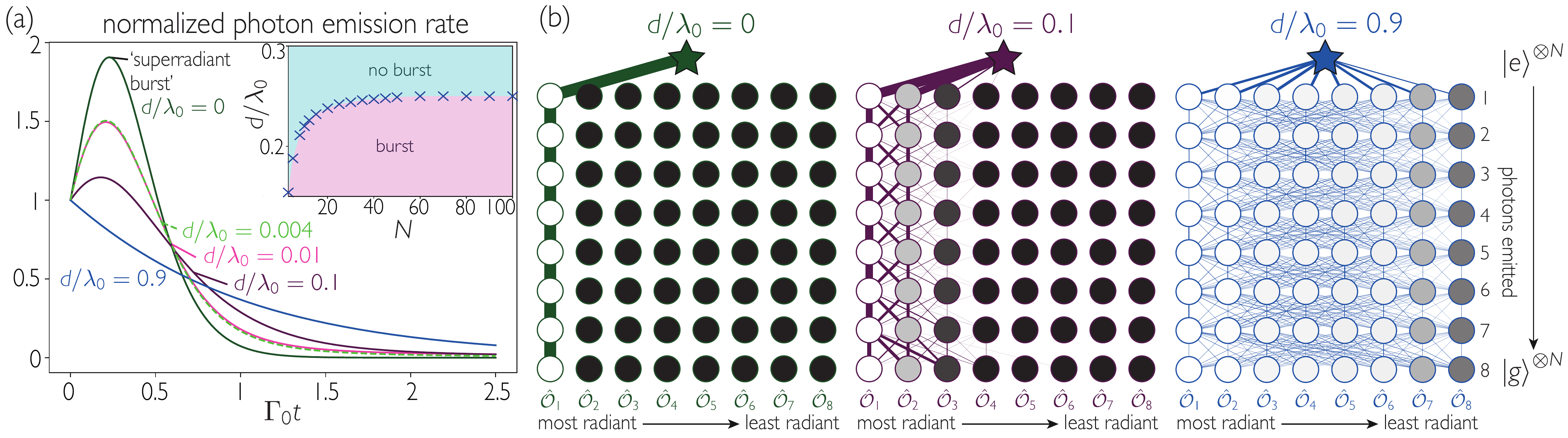}}
\caption{\textbf{Decay of a fully inverted chain and stochastic action of collective jump operators.} \textbf{(a)}~Photon emission rate normalized by atom number [$-\mathrm{d}\braket{n_{exc}(t)}/{\mathrm{d}(N\Gamma_0 t)}$] of an array of $N=8$ atoms for different interatomic distances. As $d$ is increased, the emission rate at early times shows a transition from an increase (``superradiant burst'') to a decrease. Inset: Boundary between regions. We estimate a burst  occurs if the emission rate is larger at $N\Gamma_0t=10^{-4}$ than at $t=0$. \textbf{(b)}~Illustrations of jump operator action during the decay, for different lattice constants. The star represents the initial state $\ket{e}^{\otimes N}$. Circles represent action of one of the $N$ different jump operators $\jop_{\nu}$, colored and displayed in order from most superradiant (white) on the left to most subradiant (black) on the right. Line thickness represents the likelihood of a particular path, based on a set of 1000 trajectories. For $d=0.1\lambda_0$, some trajectories are extremely subradiant (23 trajectories not fully deexcited by $\Gamma_0t = 500$ are omitted).\label{Fig2}}
\end{figure*}
   
When a jump occurs, a photon is emitted. We calculate the emission angle of the radiated photons by means of directed-detection operators, following Carmichael and coworkers~\cite{Carmichael00,Clemens03,Clemens04}. Photon detection at a point $\mathbf{R} = (r,\theta,\phi)$ corresponds to action of the operator
\begin{equation}
\hat{\mathcal{D}}(\theta,\phi) = \sqrt{\frac{3\Gamma_0}{8\pi} \left( 1 - 
\frac{\sin^2\theta}{2}\right)\mathrm{d}\Omega} \sum\limits_{j=1}^N \mathrm{e}^{-ik_0z_j\cos\theta} \hat{\sigma}^j_{ge},
\end{equation}
where $\mathrm{d}\Omega$ is a solid-angle differential. The detectors are assumed to be in far field, such that $|\mathbf{R}|\gg\lambda_0,Nd$. The probability of a photon detection in direction $(\theta,\phi)$ is $P(\theta,\phi) \mathrm{d}\Omega = \braket{\hat{\mathcal{D}}^\dagger (\theta,\phi)\hat{\mathcal{D}} (\theta,\phi)}$. A (square) photon detector of finite solid angle $\Delta\Omega$ and angular width $\Delta\theta$ sees intensity
\begin{align}
\mathfrak{I}(\theta) = \frac{\Delta\Omega}{\Delta\theta}\int\limits_{\theta-\Delta\theta/2}^{\theta+\Delta\theta/2} P(\theta') \;\mathrm{d}\theta'.
\end{align}

Photon emission caused by action of a jump operator is directional. Figure~\ref{Fig1}(b) shows the angular distribution of a photon emitted during the action of the most subradiant operator on the fully inverted array. The maximal emission angles, calculated by considering correlations between jump and directed-detection operators, are~[see Supplemental Material (SM)~\cite{SIsuperradiance}]
\nocite{Ostermann14,Schlosser02,Grunzweig10,Fung15,Brown19} 
\begin{equation}
\theta_{\mathrm{max.}} = \mathrm{arccos}\left(\frac{n\lambda_0 }{d} \pm \frac{k_\nu}{k_0}\right),\;\; n \in \mathbb{Z}.
\end{equation} 
Here, the $\pm$ accounts for the mirror reflection symmetry of a finite chain (whose jump operators carry $\pm k_{\nu}$ wave vector components). Jump operators cannot be expressed in terms of directed-detection operators~\cite{Clemens03}. For multiple holes and excitations, the intensity pattern may contain additional lobes due to atomic correlations.

A fully inverted array develops correlations as it decays. The rate of change of the atomic population, $\braket{\nexc} = \sum\nolimits_{i=1}^{N} \braket{\hat{\sigma}_{ee}^i}$, dictates the photon emission rate
\begin{equation}
R(t) = -\frac{\mathrm{d}\bnexc}{\mathrm{d}t} = \sum\limits_{\nu=1}^N \Gamma_\nu \braket{\jop^\dagger_\nu \jop_\nu}. \label{dnexcdt}
\end{equation}
At $t=0$, all atoms are excited and there are not any correlations between them ($\braket{\jop^\dagger_\nu \jop_\nu}=1\,\,\forall\, \,\nu$). By definition, $\sum_\nu \Gamma_\nu = \mathrm{Tr}~\mathbf{\Gamma} = N\Gamma_0$, and $R(t=0) = N\Gamma_0$. Since the atoms are uncorrelated, the initial decay is the sum of $N$ independently decaying atoms.

Dicke superradiance $(d=0)$ is a unique situation, as there is only one jump operator with  non-zero decay rate, $\jop_D$. That operator has rate $N\Gamma_0$ and acts identically on all atoms, i.e., $\jop_D = \left(1/\sqrt{N}\right)\sum\nolimits_{i=1}^{N}\hat{\sigma}_{ge}^i$. The decay rate $R_D(t) = N\Gamma_0\braket{\jop_D^\dagger\jop_D}$ is maximized with half the atoms excited. This gives rise to the superradiant burst seen in Fig.~\ref{Fig2}(a), where the peak rate of photon emission scales as $N^2$ and occurs at some finite time~\cite{Gross82}. Since there is a single jump operator, decay is never subradiant. Dicke superradiance seems not to be recovered as $d\rightarrow0$ [see Fig. 2(a) and the SM~\cite{SIsuperradiance}]: the emission rate saturates to a different asymptotic curve, suggesting that Dicke superradiance is not analytically connected to this regime. Rings show similar behavior~\cite{SIsuperradiance}. We attribute this saturation to a complex interplay of stimulation and competition between the set of jump operators that deexcites the array, as discussed below.

\begin{figure*}[t!]
    \centerline{\includegraphics[width=\linewidth]{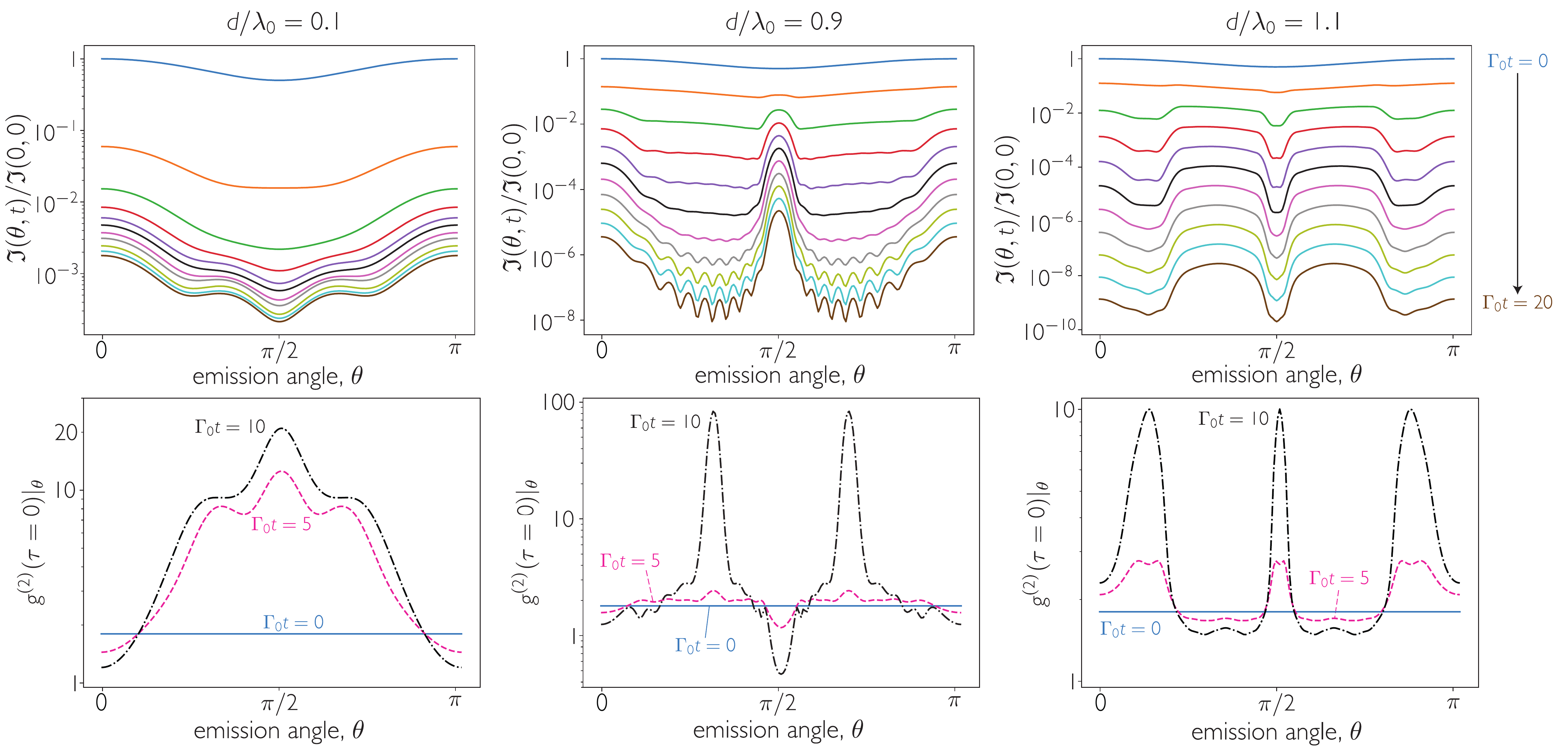}}
    \caption{\textbf{Directional photon emission from a fully inverted chain of $N=10$ atoms.} \textbf{(top)} Intensity, normalized by value at $t=0$, at far-field detectors of angular width $\Delta\theta = 0.01\pi$. Curves represent evenly spaced snapshots of the intensity profile for $\Gamma_0t \in [0, 20]$. \textbf{(bottom)} Directional two-photon correlation function, $g^{(2)}(\tau=0)|_{\theta,\theta}$ as defined in Eq.~\eqref{g2}.
\label{Fig3}}
\end{figure*}

In extended arrays, jump operators enhance their own action~\cite{Dicke54,Carmichael00,Clemens03}, but compete with each other. This occurs due to correlations induced by photon emission, irrespective of whether the photon is detected or not. For the fully inverted array, the normalized probability of two different successive jumps ($\jop_\nu$ and $\jop_\mu$) can be calculated analytically. In the large $N$ limit, it yields~\cite{SIsuperradiance}
\begin{equation}
\tilde{g}^{(2)}(\tau=0)|_{\nu,\mu} = \frac{\braket{\jop^\dagger_\nu \jop^\dagger_\mu \jop_\mu \jop_\nu}}{\braket{\jop^\dagger_\nu\jop_\nu}\braket{\jop^\dagger_\mu\jop_\mu}} \simeq 1 + \delta_{\nu\mu} - \frac{2}{N}.
\end{equation}
While this is a process of spontaneous emission, each jump operator enhances itself (but not others), and thus an effective stimulated emission of radiation occurs at certain angles, as jump operators are directional. The last term in the equation is a fermionic correction that illustrates that a single atom can only host a single excitation (there is a $1/N$ probability for two excitations to overlap in a chain of $N$ atoms, and the factor of 2 arises because there are two identical ways to assign two excitations). For $N \geq 4$, $\tilde{g}^{(2)}(\tau=0)|_{\nu,\nu} > 1~\forall~\nu$, i.e., even in very short chains all operators enhance their own action.

The competition between different jump operators causes dephasing of the atomic state, reducing and eventually blocking superradiance. Following this argument, atoms in other geometries must also dephase, since for $d\neq0$ there are always multiple operators and, thus, competition between decay paths. The set of likely deexcitation paths diversifies as $d$ increases, as shown in Fig.~\ref{Fig2}(b), leading to faster dephasing. This reduces the intensity of the superradiant burst and brings it forward in time - no longer happening when half of the atoms are excited, but earlier - until the burst disappears. Each path has a likelihood dictated by the operator decay rates and the correlations induced as the ensemble deexcites. The only forbidden paths are those where the cumulative effect of the jump operators breaks the mirror symmetry about the center of the array. 

We find that the superradiant burst survives at small enough interatomic distances despite being suppressed, as shown in Fig.~\ref{Fig2}(a). While the peak intensity is fainter and does not scale as $N^2$~\cite{SIsuperradiance}, the photon emission rate initially increases. In the inset, we show the distance for which the superradiant burst disappears. We estimate that a superradiant burst will occur if the derivative of the photon emission rate at $t=0$ is positive. We perform calculations for large atom numbers by truncating the Hilbert space to subspaces with up to two atoms in the ground state (i.e., maximum of two photons emitted). This captures well the early dynamics. As shown in the inset of Fig.~\ref{Fig2}(a), for long chains, superradiant features are retained for $d\lesssim 0.25\lambda_0$. At this upper limit, competition between jump operators becomes so strong that the burst becomes marginal and is eventually suppressed.

The action of different jump operators throughout the evolution leads to changes in the directionality of photon emission at different times, as shown in Fig.~\ref{Fig3}. The fully excited ensemble emits quite broadly in space. Without correlations, the atoms emit as independently radiating dipoles~\cite{Carmichael00,Clemens03}. However, at late times, subradiant operators become dominant and emission is strongly peaked in a direction dictated by $d$. Angular emission is narrow for $d=0.9\lambda_0$, as there is one dominant subradiant mode, but broad for $d=1.1\lambda_0$ where multiple subradiant modes are important [see Fig.~\ref{Fig1}(b)]. Radiation in different directions is correlated~\cite{Liberal19}: emission at angle $\theta_1$ enhances emission in directions that satisfy $\cos\theta_2 = \cos\theta_1 - n\lambda_0/d,\;n\in\mathbb{Z}$, as jump operator emission patterns are multilobed~\cite{SIsuperradiance}. 

Self-enhancing (or stimulated) emission is confirmed by calculating the direction-dependent second order correlation function
\begin{equation}
g^{(2)}(\tau=0)|_{\theta,\theta} = \frac{\braket{\dop^\dagger(\theta)\dop^\dagger(\theta)\dop(\theta)\dop(\theta)}}{\braket{\dop^\dagger(\theta)\dop(\theta)}^2}.\label{g2}
\end{equation}
Figure~\ref{Fig3} shows large, direction-dependent bunching in the field radiated by the array under evolution according to Eq.~(\ref{mejumps}). At $t=0$, $g^{(2)}(\tau=0)|_{\theta,\theta}$ can be calculated analytically, yielding a spatially uniform value of $2 - 2/N$~\cite{SIsuperradiance}, reproducing Dicke's result~\cite{Dicke54,Clemens04}. At late times, there are large peaks at intensity minima. While single photon emission is very unlikely, conditioned on one photon, a second is significantly more likely, such that pairs are relatively enhanced~\cite{Bhatti15,Gulfam18}. At late times, the signal can be sub-Poissonian in the direction of peak intensity [see plot for $d=0.9\lambda_0$ in Fig.~\ref{Fig3}], as subradiance is predominantly a single-excitation effect and photon pairs are suppressed. Evidence of such directional statistics has been observed for two emitters~\cite{Wolf20}.

\begin{figure}
\centerline{\includegraphics[width=\linewidth]{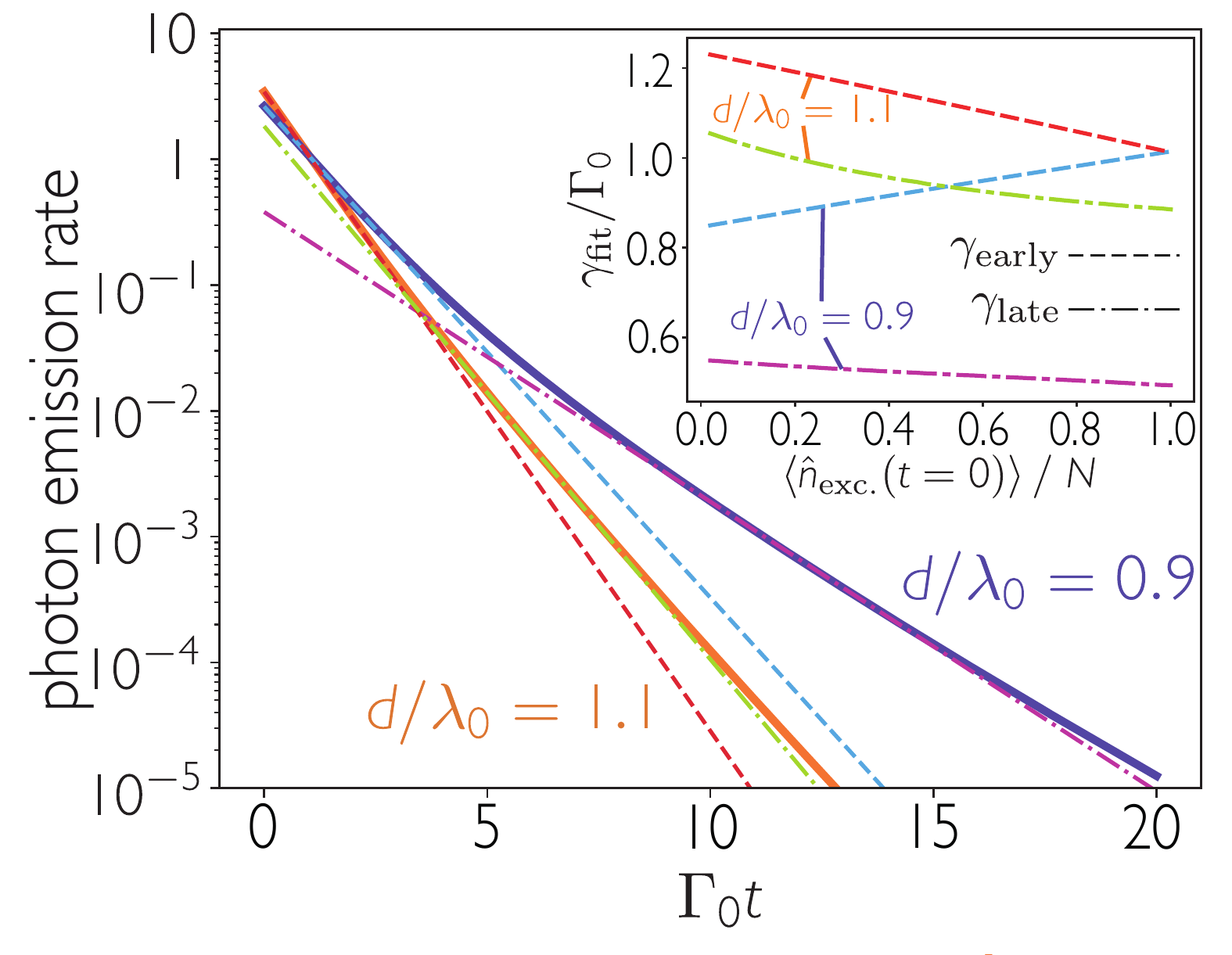}}
\caption{\textbf{Role of initial number of excitations on photon emission rate from spin coherent states.} Photon emission (solid lines) from $\ket{\varphi=0.3,\mathbf{k}=k_0\hat{z}}$ with average excitation number $\braket{\nexc}= \varphi N$ [see Eq.~\eqref{coherenteqn}]. Dashed lines show exponential fits, from which one extracts decay rate coefficients $\gamma_\text{early}$ (dashed) and $\gamma_\text{late}$ (dot-dashed). Inset: Early and late fitted decay rates as a function of the initial number of excitations. For all plots, $N=10$. For $d=0.9\lambda_0$, $\gamma_{\mathrm{early}}$ is fitted for $\Gamma_0t \in [0,4]$ and $\gamma_{\mathrm{late}}$ for $\Gamma_0t \in [10,20]$. For $d=1.1\lambda_0$, $\gamma_{\mathrm{early}}$ is fitted for $\Gamma_0t \in [0,2]$ and $\gamma_{\mathrm{late}}$ for $\Gamma_0t \in [5,10]$.\label{Fig4}}
\end{figure}

Signatures of collective decay can be observed without fully inverting the array, but preparing spin coherent states, instead~\cite{Arecchi72},
\begin{equation}
\ket{\varphi,\mathbf{k}} = \prod\limits_{i=1}^N \sqrt{1-\varphi} \ket{g_i} + \mathrm{e}^{i\mathbf{k}\cdot\rb_i}\sqrt{\varphi} \ket{e_i},\label{coherenteqn}
\end{equation}
where $0 \leq \varphi \leq 1$. These states have a binomially distributed excitation number defined by probability $\varphi$, and an excitation expectation value $\braket{\nexc} = \varphi N$. Coherent spin states can be prepared experimentally by exciting the array with an intense pulse of duration $\tau \ll (N\Gamma_0)^{-1}, (J^{12})^{-1}$, to prevent collective effects.
    
Coherent spin states exhibit nonexponential temporal decay due to the interplay of multiple jump operators. The subradiant tail survives at distances accessible with current experimental capabilities, as shown in Fig.~\ref{Fig4}. This can be characterized by separately fitting the early and late dynamics, as demonstrated in  experiments with atomic clouds in free space~\cite{Guerin16,Cipris20arxiv} and near a nanofiber~\cite{Solano17}. The fitted decay rates of early dynamics depend on the initial atomic state; the decay is a many-body problem dependent on the density of excitations. Fits at late times do not depend so markedly on the number of excitations, since late dynamics are predominantly a single-excitation phenomenon independent of initial conditions. For large numbers of excitations, $\gamma_{\mathrm{early}}$ converges for all geometries as the fully excited state is uncorrelated. The contrast between early and late fits is significantly larger for $d = n\lambda_0/2 - \epsilon$, due to the differences in decay rates across each resonance. As shown in the SM~\cite{SIsuperradiance}, the drive can be used to imprint correlations on the array, some of which survive at long times and impact the radiation pattern.

Experimental realizations in the regime $d\sim\lambda_0$ should be feasible with current technologies. However, reaching the superradiant regime requires shorter interparticle distances. The limit of $d\lesssim0.25\lambda_0$ established here could be satisfied with long-wavelength transitions (such that $\lambda_0>800$~nm) of lanthanide atoms trapped in optical lattices with wavelength near their strong transitions $\sim400$~nm~\cite{Lepers14,Li16}. Even shorter interatomic distances can be reached in disordered ensembles~\cite{Pellegrino14,Corman17}, which constitute an interesting prospect for future work. Arrays of solid state emitters, such as localized excitonic quantum dots or strain-generated defects in 2D materials~\cite{Palacios17,Proscia18}, are an additional playground for collective decay.

In conclusion, we have studied the collective decay of ordered chains of atoms in free space. We have found that superradiance survives significant interatomic separations, though Dicke's  perfectly symmetric decay is lost at any finite distance in all ordered geometries. For separations comparable to the resonance wavelength, strong signatures of collective decay remain and photon emission has directional features and a subradiant tail. These phenomena are robust to realistic experimental imperfections, namely finite filling fraction and classical position noise~\cite{SIsuperradiance}. Geometry can be used to control the many-body optical response of arrays. With conditional feedback control~\cite{Smith02} (assisted by directional detection), it may pave the way toward the preparation of target entangled states, such as metrologically-useful subradiant states.

Research on superradiance and many-body physics was supported by Programmable Quantum Materials, an Energy Frontier Research Center funded by the U.S. Department of Energy (DOE), Office of Science, Basic Energy Sciences (BES), under Award No. DE-SC0019443. Research on late-time dynamics and subradiance was supported by the National Science Foundation QII-TAQS (Grant No. 1936359). I.F.B. and A.B. acknowledge financial support by the R\'{e}gion \^{I}le-de-France in the framework of Domaine d'Int\'{e}r\^{e}t Majeur SIRTEQ (Project DSHAPE).

\clearpage
\onecolumngrid

\section{Supplemental Material}

This Supplemental Material contains derivations of correlation functions of decay operators for a fully-inverted 1D atomic array, further details on the limits of superradiance, properties of non-exponential decay for experimentally-accessible states, and a discussion on the role of imperfections in the collective decay of 1D arrays.

\section{Contents}

\noindent 1\;\;\;Correlations between jump operators \hfill 8\\

1.1\;\;\;Successive operator action: Infinite chain \hfill 8\\

1.2\;\;\;Successive operator action: Finite chain \hfill 9\\

1.3\;\;\;Multi-operator action\hfill 10\\

1.4\;\;\;Directional correlations\hfill 11\\

1.5\;\;\;Directional emission of jump operators\hfill 11\\

\noindent 2\;\;\;Scalings of the superradiant burst\hfill 12\\

2.1\;\;\;Short distances $d/\lambda_0\rightarrow 0$ versus Dicke limit $d=0$ \hfill 12\\

2.2\;\;\;Scaling of maximum operator decay rate \hfill 13\\

2.3\;\;\;Scaling of subradiance \hfill 13\\

\noindent 3\;\;\;Non-exponential decay with different initial states\hfill 14\\

3.1\;\;\;Coherent spin states \hfill 14\\

3.2\;\;\;Collectively driven states \hfill 15\\

3.3\;\;\;``Memory'' of the initial state \hfill 16\\

\noindent 4\;\;\;Role of imperfections \hfill 17

\subsection{1\;\;\;Correlations between jump operators}
Here, we demonstrate correlations between jump operator action. We analyze both infinite and finite chains, making calculations for the fully-inverted array. This allows us to derive analytical expressions for correlations. We demonstrate that each jump operator enhances its own action.

\subsubsection{1.1\;\;\;Successive operator action: Infinite chain}

The normalized probability of two different successive jumps for jump operators $\jop_\nu, \jop_\mu$ is
\begin{equation}
\tilde{g}^{(2)}(\tau=0)|_{\nu,\mu} = \frac{\braket{\jop_\nu^\dagger \jop_\mu^\dagger \jop_\mu \jop_\nu}}{\braket{\jop_\nu^\dagger \jop_\nu}\braket{\jop_\mu^\dagger \jop_\mu}}.
\end{equation}

For an infinite system, the jump operators are Bloch waves, $\jop_\nu = \left(1/\sqrt{N}\right)\sum_{i=1}^N \mathrm{e}^{-ik_\nu z_i}\hge^i$. This allows us to analytically study these probabilities. For $\nu=\mu$,
\begin{align}
\tilde{g}^{(2)}(\tau=0)|_{\nu,\nu} &= \frac{\braket{\jop^\dagger_\nu\jop^\dagger_\nu\jop_\nu\jop_\nu}}{\braket{\jop^\dagger_\nu\jop_\nu}^2}= \frac{\sum\limits_{i=1}^N \sum\limits_{j=1}^N \sum\limits_{l=1}^N \sum\limits_{m=1}^N \mathrm{e}^{ik_\nu(z_i + z_j - z_l - z_m)} \braket{\heg^i \heg^j \hge^l \hge^m}}{\left(\sum\limits_{i=1}^N \sum\limits_{l=1}^N \mathrm{e}^{ik_\nu(z_i - z_l)}\braket{\heg^i\hge^l}\right)^2}.
\end{align}
We evaluate this quantity on the fully-inverted state $\ket{e}^{\otimes N}$. The only non-zero contributions to the sums are those where the operators act to return the atoms to a fully-inverted state, i.e., those where the indices of the lowering and raising operators are the same. This thus yields
\begin{align}
\tilde{g}^{(2)}(\tau=0)|_{\nu,\nu} &= \frac{2\sum\limits_{i=1}^N \sum\limits_{j=1}^N \braket{\heg^i \heg^j \hge^i \hge^j}}{\left(\sum\limits_{i=1}^N \braket{\heg^i\hge^i}\right)^2}= 2 - \frac{2}{N} \xrightarrow{N\rightarrow\infty} 2.
\end{align}
The factor of two arises from the two equivalent combinations of $\{i,j\}=\{l,m\}$, and is mentioned in Dicke's original superradiance paper~\cite{Dicke54}. The correction $2/N$ accounts for terms in the sum where $i=j$, and is a result of the two-level nature of the atoms, which we term the fermionic correction. Note that there is no dependence on the geometry of the array, beyond the Bloch wave assumption.

For $\mu \neq \nu$, the probabilities are calculated as
\begin{align}
\tilde{g}^{(2)}(\tau=0)|_{\nu,\mu} &= \frac{\sum\limits_{i=1}^N \sum\limits_{j=1}^N \sum\limits_{l=1}^N \sum\limits_{m=1}^N \mathrm{e}^{ik_\nu(z_i - z_l)}\mathrm{e}^{ik_\mu(z_j - z_m)} \braket{\heg^i \heg^j \hge^l \hge^m}}{\left(\sum\limits_{i=1}^N \sum\limits_{l=1}^N \mathrm{e}^{ik_\nu(z_i - z_l)}\braket{\heg^i\hge^l}\right)\left(\sum\limits_{j=1}^N \sum\limits_{m=1}^N \mathrm{e}^{ik_\mu(z_j - z_m)}\braket{\heg^j\hge^m}\right)}.
\end{align}
In this case, the two combinations of $\{i,j\}=\{l,m\}$ are not equivalent. This leads to
\begin{align}\nonumber
\tilde{g}^{(2)}(\tau=0)|_{\nu,\mu} &= \frac{\sum\limits_{i=1}^N \sum\limits_{j=1}^N \left(1 + \mathrm{e}^{i(k_\nu-k_\mu)(z_i-z_j)}\right)\braket{\heg^i\heg^j\hge^i\hge^j}}{\left(\sum\limits_{i=1}^N \braket{\heg^i\hge^i}\right)\left(\sum\limits_{j=1}^N \braket{\heg^j\hge^j}\right)}\\
&= 1 - \frac{2}{N} +\frac{1}{N^2} \sum\limits_{i=1}^N\sum\limits_{j=1}^{N}\mathrm{e}^{i(k_\nu-k_\mu)(z_i-z_j)} \xrightarrow{N\gg 1} 1 + \delta_{\nu\mu}- \frac{2}{N}.
\end{align}
In the infinite limit, the fermionic correction becomes negligible, such that operators enhance themselves and have no correlations with other operators.

\subsubsection{1.2\;\;\;Successive operator action: Finite chain}

For a finite array, the jump operators are mirror symmetric about the center of the array and are represented by sums over $\pm k_\nu$. We can analytically study the normalized probabilities using the ansatz~\cite{Asenjo17PRX}
\begin{subequations}
\begin{align}
\jop_\nu &= \sqrt{\frac{2}{N+1}} \sum\limits_{i=1}^N \sin(k_\nu z_i) \hge^i\;\;\mathrm{if}~\nu~\mathrm{even}, \\
\jop_\nu &= \sqrt{\frac{2}{N+1}} \sum\limits_{i=1}^N \cos(k_\nu z_i) \hge^i\;\;\mathrm{if}~\nu~\mathrm{odd}.
\end{align}
\end{subequations}
Note that here we diagonalize $\mathbf{\Gamma}$, rather than the effective Hamiltonian~\cite{Carmichael00,Clemens03}. From now on, we consider even $\nu$, but all equations hold for odd $\nu$ with the prescription $\sin(\cdot)\rightarrow\cos(\cdot)$.

For $\nu=\mu$ and even $\nu$:
\begin{align}
\tilde{g}^{(2)}(\tau=0)|_{\nu,\nu} &= \frac{\braket{\jop^\dagger_\nu\jop^\dagger_\nu\jop_\nu\jop_\nu}}{\braket{\jop^\dagger_\nu\jop_\nu}^2}= \frac{\sum\limits_{i=1}^N \sum\limits_{j=1}^N \sum\limits_{l=1}^N \sum\limits_{m=1}^N \ \sin(k_\nu z_i)\sin(k_\nu z_j) \sin(k_\nu z_l) \sin(k_\nu z_m) \braket{\heg^i \heg^j \hge^l \hge^m}}{\left(\sum\limits_{i=1}^N \sum\limits_{l=1}^N \sin(k_\nu z_i) \sin(k_\nu z_l) \braket{\heg^i\hge^l}\right)^2}.
\end{align}
As before, we evaluate this quantity on the fully-inverted state. This allows the simplification
\begin{align}
\tilde{g}^{(2)}(\tau=0)|_{\nu,\nu} &= \frac{2\sum\limits_{i=1}^N \sum\limits_{j=1}^N \sin^2(k_\nu z_i) \sin^2(k_\nu z_j)\braket{\heg^i \heg^j \hge^i \hge^j}}{\left(\sum\limits_{i=1}^N \sin^2(k_\nu z_i)\braket{\heg^i\hge^i}\right)^2}= 2 - \frac{2\sum\limits_{i=1}^N \sin^4(k_\nu z_i)}{\left(\sum\limits_{i=1}^N \sin^2(k_\nu z_i) \right)^2}.
\end{align}
In comparison to the Bloch wave calculation, the fermionic correction has a more complicated form. However, in the large $N$ limit, the same result $\tilde{g}^{(2)}(\tau=0)|_{\nu,\nu}=2$ is reached.

For $\mu \neq \nu$ and even $\nu,\mu$, we calculate the probabilities as
\begin{align}
\tilde{g}^{(2)}(\tau=0)|_{\nu,\mu} &= \frac{\sum\limits_{i=1}^N\sum\limits_{j=1}^N\sum\limits_{l=1}^N\sum\limits_{m=1}^N \sin(k_\nu z_i)\sin(k_\mu z_j) \sin(k_\mu z_l) \sin(k_\nu z_m) \braket{\heg^i\heg^j\hge^l\hge^m}}{\left(\sum\limits_{i=1}^N\sum\limits_{l=1}^N \sin(k_\nu z_i)\sin(k_\nu z_l)\braket{\heg^i\hge^l}\right)\left(\sum\limits_{i=1}^N\sum\limits_{l=1}^N \sin(k_\mu z_i)\sin(k_\mu z_l) \braket{\heg^i\hge^l}\right)} \\
&= \frac{\sum\limits_{i=1}^N\sum\limits_{j=1}^N \left[ \sin^2(k_\nu z_i)\sin^2(k_\mu z_j) + \sin(k_\nu z_i)\sin(k_\mu z_i) \sin(k_\mu z_j)\sin(k_\nu z_j)\right]\braket{\heg^i\heg^j\hge^j\hge^i}}{\left(\sum\limits_{i=1}^N \sin^2(k_\nu z_i)\braket{\heg^i\hge^i}\right)\left(\sum\limits_{i=1}^N\sin^2(k_\mu z_i) \braket{\heg^i\hge^i}\right)} \\
&= 1 + \frac{\sum\limits_{i=1}^N\sum\limits_{j=1}^N \sin(k_\nu z_i)\sin(k_\mu z_i) \sin(k_\mu z_j)\sin(k_\nu z_j)}{\left(\sum\limits_{i=1}^N \sin^2(k_\nu z_i) \right)\left(\sum\limits_{i=1}^N\sin^2(k_\mu z_i) \right)} - \frac{2\sum\limits_{i=1}^N \sin^2(k_\nu z_i)\sin^2(k_\mu z_i)}{\left(\sum\limits_{i=1}^N \sin^2(k_\nu z_i) \right)\left(\sum\limits_{i=1}^N\sin^2(k_\mu z_i) \right)}.
\end{align}
The fermionic correction again takes a more complicated form than in the Bloch wave picture. In addition, the correlation term is also more complicated. However, the same results can be inferred. The correlation sum is maximized for $\nu=\mu$ and must always have value less than this for $\nu \neq \mu$. As in the infinite case, operators enhance their own action more than that of any other operator.

\subsubsection{1.3\;\;\;Multi-operator action}

We can calculate normalized probabilities for the action of a single operator for an infinite chain. For $M\geq 2$ actions by one operator, this is given by
\begin{equation}
\tilde{g}^{(M)}(\tau=0)|_{\nu,\nu} = \frac{\left\langle\left(\jop^\dagger_\nu\right)^M \left(\jop_\nu\right)^M\right\rangle}{\braket{\jop^\dagger_\nu\jop_\nu}^M}. 
\end{equation}
The denominator simplifies to $N^M$. The numerator has a leading term $N^M$ multiplied by the number of available ways of arranging the indices of the raising operators to coincide with those of the lowering operators, $M!$. The corrections are then the number of terms in the sum where two or more indices of the lowering operators are the same. We can split corrections into terms with $S$ matching indices. Each correction term is the size of the set of terms with $S$ matching indices, multiplied by the number of times that term appears in the rest of the sum. This results in the formula
\begin{align}
\tilde{g}^{(M)}(\tau=0)|_\nu &= M! - \frac{1}{N^M}\sum\limits_{S=2}^M \binom{M}{S} N (N-1)^{M-S}.
\end{align}
where $\binom{M}{S}$ is a binomial coefficient. Notably, in the infinite $N$ limit, the $M$-photon correlation function is $M!$, meaning that operator action is further and further enhanced as the atoms lock phase. We note that such a formula has limitations, since it calculates statistics only at $t=0$, and does not capture temporal evolution of the state (i.e., it does not include Hamiltonian evolution or action of a different jump operator).

\subsubsection{1.4\;\;\;Directional correlations}

For directional operators acting on the fully-inverted array, the second order correlation function yields the same result as that for the jump operators, i.e.,  
\begin{align}
g^{(2)}(\tau=0)|_{\theta,\theta} &= \frac{\braket{\dop^\dagger(\theta)\dop^\dagger(\theta)\dop(\theta)\dop(\theta)}}{\braket{\dop^\dagger(\theta)\dop(\theta)}^2}= 2 - \frac{2}{N}.
\end{align}
The directional operators are characterized by their emission wave-vector, and produce identical results for both finite and infinite lattices. The values of $g^{(2)}(\tau=0)|_{\theta,\theta}$ are verified by numerical plots such as those in Fig.~3 of the main text. 

\begin{figure}[b]
    \centering
    \includegraphics[width=.95\textwidth]{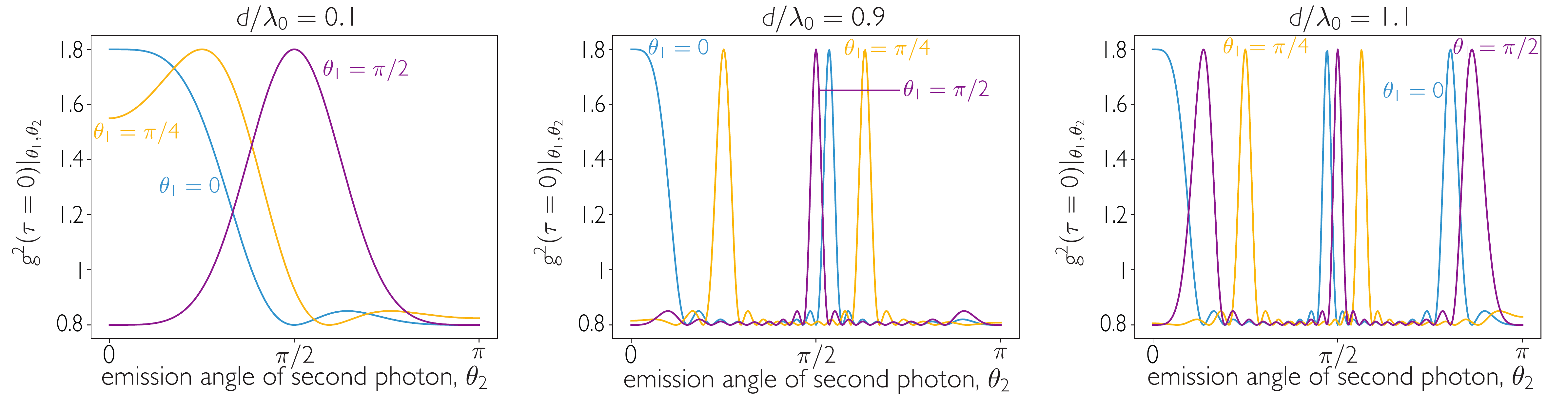}
    \caption{Directional cross-correlations at different lattice spacings. $g^{(2)}(\tau=0)|_{\theta_1,\theta_2}$, as defined in Eq.~\eqref{g2t12}, for a fully-inverted array of $N=10$ atoms polarized perpendicular to the chain axis.}
    \label{SIFig7}
\end{figure}

For two different directional operators, the correlation function is
\begin{equation}\label{g2t12}
g^{(2)}(\tau=0)|_{\theta_1,\theta_2} = 1 - \frac{1}{N} +\frac{1}{N^2} \sum\limits_{i=1}^N\sum\limits_{j=1, j\neq i}^{N}\mathrm{e}^{ik_0(\cos\theta_1-\cos\theta_2)(z_i-z_j)}.
\end{equation}
Different directional operators enhance each other's action if $(\cos\theta_1 - \cos\theta_2)d/\lambda_0 = n,\;n\in\mathbb{Z}$. For $d< \lambda_0/2$ only the $n=0$ solution exists, which gives $\theta_1 = \theta_2$. At larger distances, the emission of a photon at $\theta_1$ can enhance photon emission in multiple directions. This is shown in Fig.~\ref{SIFig7}. Primary lobes peaked at $\theta_1$ appear at all distances, but for $d=0.9\lambda_0$ and $d=1.1\lambda_0$ correlated peaks in other directions appear with equal magnitude.

\subsubsection{1.5\;\;\;Directional emission of jump operators}

We can also consider single-photon correlations between the jump operators and directional operators, when acting on a fully-inverted chain. For a finite chain and even $\nu$ 
\begin{align}\nonumber
D(\nu,\theta) &= \braket{\dop^\dagger(\theta) \jop_\nu}
= C(\theta) \sum\limits_{i=1}^N\sum\limits_{j=1}^N \mathrm{e}^{ik_0\cos\theta z_i}\sin(k_\nu z_j)\braket{\heg^i\hge^j} \\
&= C(\theta)\sum\limits_{i=1}^N \mathrm{e}^{ik_0\cos\theta z_i}\sin(k_\nu z_i) = \frac{C(\theta)}{2\ii} \sum\limits_{i=1}^N \left[\mathrm{e}^{i(k_0\cos\theta + k_\nu)z_i} - \mathrm{e}^{i(k_0\cos\theta - k_\nu)z_i}\right],
\end{align}
where $C(\theta)$ is a constant containing the pre-factors for $\dop^\dagger(\theta)$ and $\jop_\nu$. Peaks in the emission pattern are given by solutions to $(k_0 \cos\theta_{\mathrm{max.}} \pm k_\nu)d = 2\pi n$, with $n \in \mathbb{Z}$. For $d < \lambda_0 / 2$, there are two emission lobes (for $n=0$) at $\theta_{\mathrm{max.}} = \mathrm{arccos}\left(\pm k_\nu / k_0\right)$. As $d$ increases, solutions with higher $n$ are admitted. In particular, for distances $n\lambda_0/2<d<(n+1)\lambda_0/2$, there are a maximum of $2(n+1)$ emission lobes, exactly as in classical antenna phased arrays. These lobes may coalesce into each other. The emergence of a new lobe is correlated with the sudden jumps in decay rates. For odd $\nu$, the same angular solutions arise, as the prescription of a cosine does not alter the exponentials.

\subsection{2\;\;\;Scalings of the superradiance burst}

\subsubsection{2.1\;\;\;Short distances $d/\lambda_0\rightarrow0$ versus Dicke limit $d=0$}

\begin{figure}[b!]
\centerline{\includegraphics[width=.75\linewidth]{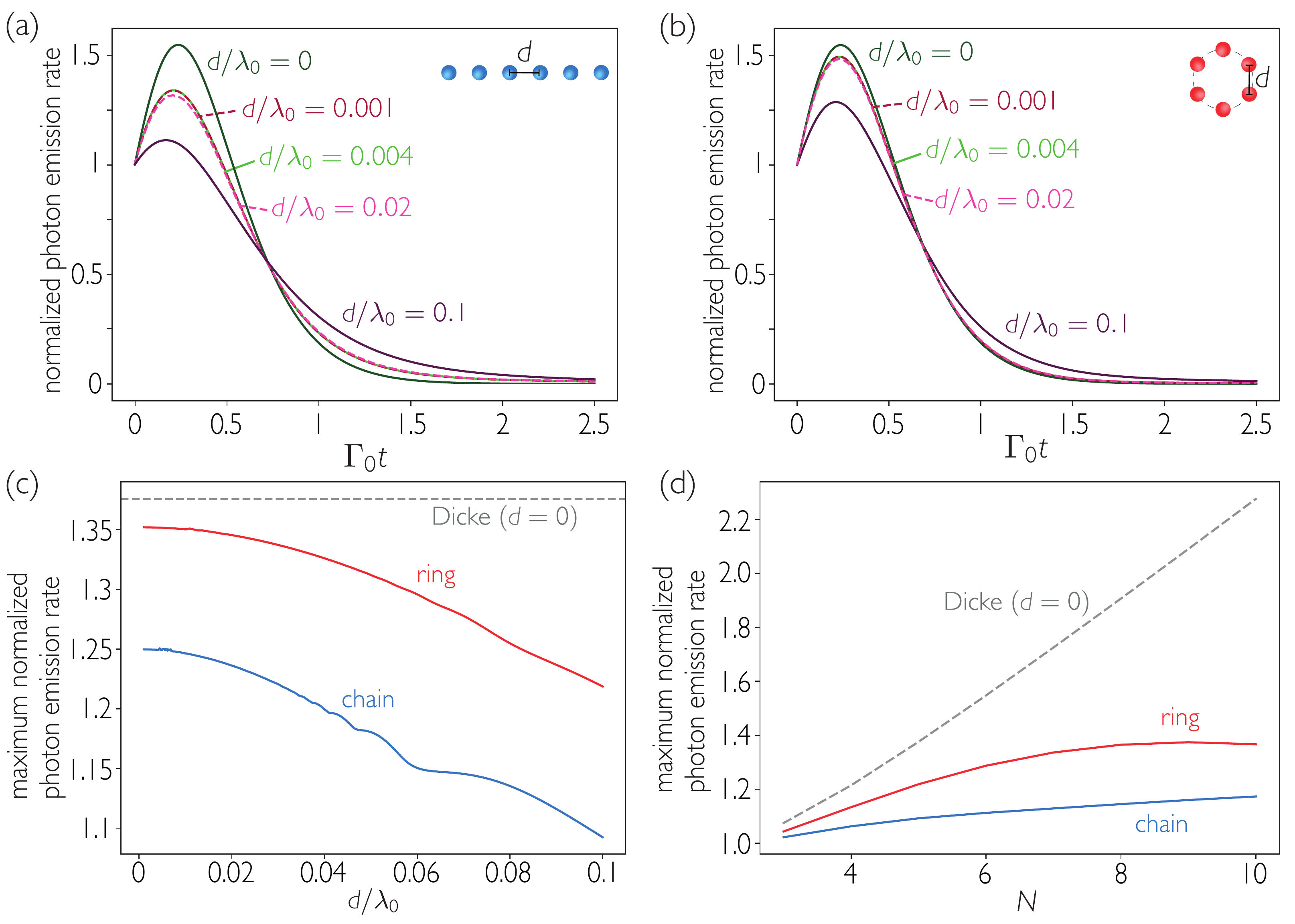}}
\caption{Superradiance from fully-inverted chains and rings. \textbf{(a,b)}~Rate of photon emission normalized by atom number [$-\mathrm{d}\braket{n_{exc}(t)}/{\mathrm{d}(N\Gamma_0 t)}$] of an array of $N=6$ atoms for different inter-atomic distances in chain and ring configurations. \textbf{(c,d)}~Peak of photon emission normalized by atom number for (c)~$N=5$ atoms as $d\rightarrow0$ and (d)~$d=0.1\lambda_0$. In all cases, atoms are polarized perpendicular to the array.\label{SIFig4}}
\end{figure}

Whether the decay of an array of atoms of finite size, characterized by an inter-atomic distance $d$, converges to the $d=0$ case as $d\rightarrow0$ has been a matter of historical debate~\cite{Gross82,BenedictBook}. In Fig.~\ref{SIFig4}, we show the emitted photon pulse at different inter-atomic distances for two different geometries, a chain and a ring. For very small $d$, reduction in the inter-atomic spacing does not seem to change the temporal profile of the emitted intensity. Either the limit is approached extremely slowly or non-monotonically, or there is a discontinuity in the evolution at $d=0$ and infinitesimally small $d$. Certainly, for any physical system in free space, where the emitters must themselves be of finite size and have strong short-range van der Waals interactions, the case of Dicke's perfectly symmetrical superradiance cannot be achieved. In addition, at very small distances, the frequency shifts of the collective modes are large enough (on the order of $10^6\Gamma_0$ for $d=0.001\lambda_0$) that the Born-Markov approximation used to eliminate the field becomes invalid. As $N$ is increased, the maximum burst intensity deviates from the $N^2$ scaling, that of the Dicke case [see Fig.~\ref{SIFig4}(d)]. This implies that the difference is not a finite size effect.

We find that superradiance in a ring geometry also deviates from the Dicke scenario. It has been claimed that superradiance should survive finite separations for particular symmetric geometries, such as rings~\cite{Gross82,BenedictBook}. This was explained by consideration of the symmetry of the atomic states. The argument is as follows: For an atomic chain, the Hamiltonian is not symmetric for atom exchange and different atoms see different environments depending on their position in the chain. As the ensemble decays, this dephases the atoms, leading to suppression of superradiance. For atoms arranged in a ring, the Hamiltonian is symmetric under atom exchange and all atoms initially see the same field, hence superradiance is predicted to survive. However, we find that this is not true. We instead attribute the dephasing to a competition between different jump operators. Jump operators compete to induce different patterns of phase-locking in any geometry and the state will dephase as it decays.

\subsubsection{2.2\;\;\;Scaling of maximum operator decay rate}

\begin{figure}[h]
    \centerline{\includegraphics[width=.75\linewidth]{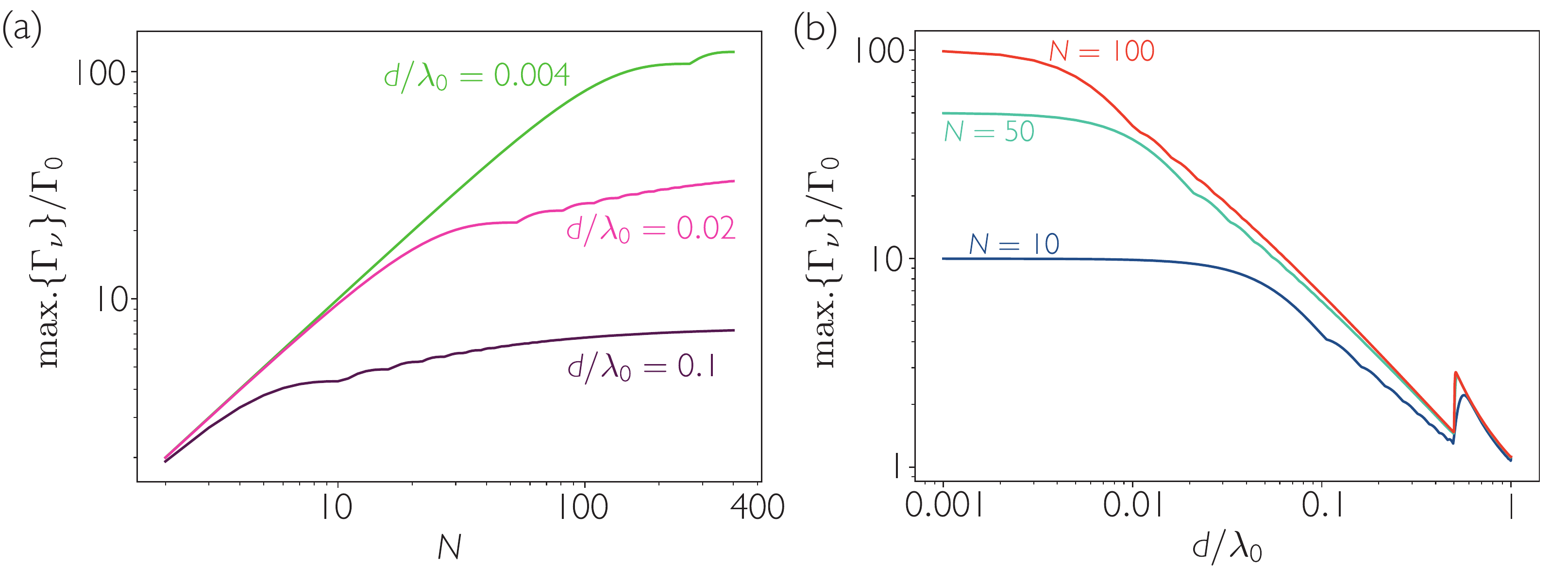}}
\caption{Largest eigenvalue of the dissipative interaction matrix $\mathbf{\Gamma}$ for a 1D array with polarization perpendicular to the array.
\label{SIFig6}}
\end{figure}

The largest eigenvalue of the dissipative interaction matrix $\mathbf{\Gamma}$ does not scale with $N$ for finite distances, in contrast to the $d=0$ case. This is because, as $N$ is increased, a second mode eventually appears inside the light cone (defined as the region $|k_z|<k_0$~\cite{Asenjo17PRX}), and the emission is split between two competing decay processes. As shown in Fig.~\ref{SIFig6}(a), this leads to linear scaling of the rate up to some value of $N$ where a second operator becomes significantly bright. At small $d$, subradiant paths are not completely dark due to finite size effects, and at large $d$ there are multiple bright paths. This means that for all $N$, there will be multiple decay paths leading to de-phasing of the superradiant burst. For fixed $N$, the rate increases to $N\Gamma_0$ as $d\rightarrow0$, as shown in Fig.~\ref{SIFig6}(b). Saturation occurs at lower values of $d$ for higher $N$, indicating that the saturation of the maximum decay rate is related to the total length of the array.

\subsubsection{2.3\;\;\;Scaling of subradiance}

\begin{figure}[b!]
    \centerline{\includegraphics[width=.75\linewidth]{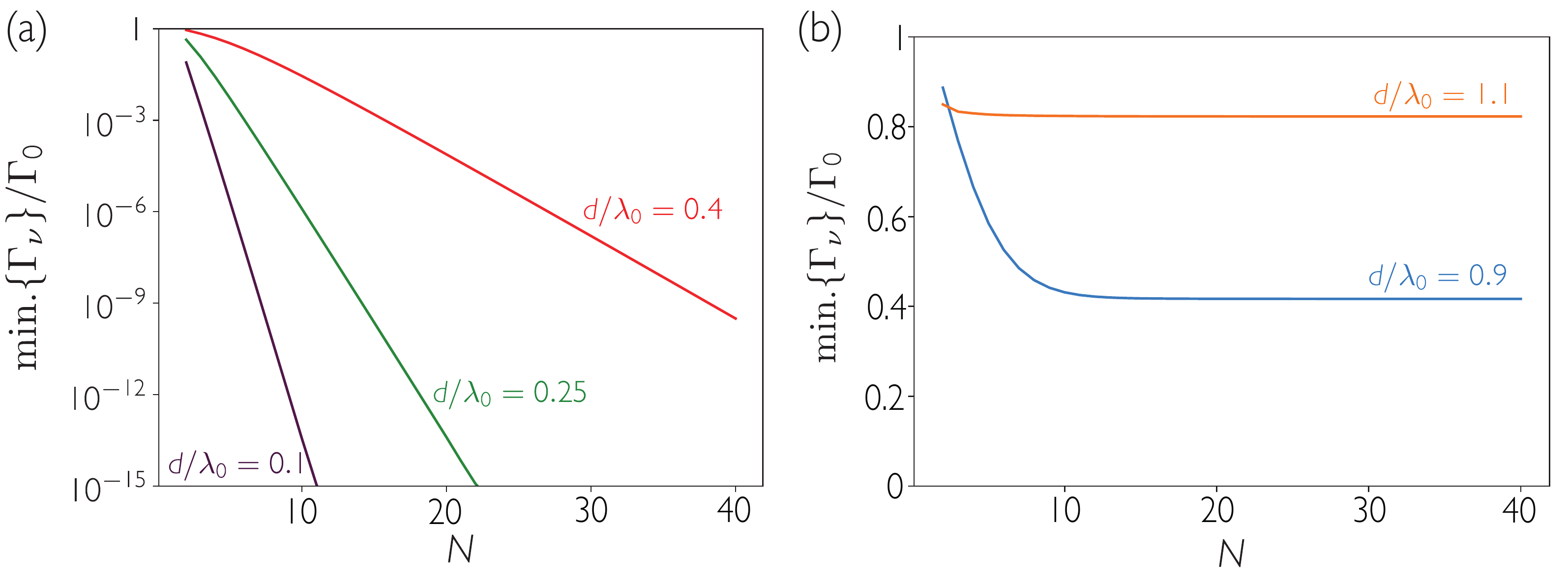}}
\caption{Smallest eigenvalue of the dissipative interaction matrix $\mathbf{\Gamma}$ for a 1D array with polarization perpendicular to the array.
\label{SIFig9}}
\end{figure}

Late dynamics of a decaying ensemble are governed by subradiance. As the array reaches single-excitation subradiant states, only the action of the subradiant operators can fully de-excite the state. The characteristic decay time of late dynamics is thus governed by the decay rates of the most subradiant operators. For $d < \lambda_0/2$, the smallest eigenvalue of the dissipative interaction matrix decreases exponentially with atom number, as shown in Fig.~\ref{SIFig9}(a). For large atom numbers, the set of rates approaches a continuum, leading to power law temporal decay of population at late times~\cite{Henriet19}. In contrast, for $d > \lambda_0/2$ the smallest eigenvalue saturates. As shown in Fig.~\ref{SIFig9}(b), this saturation occurs for small $N$. Therefore, the late dynamics of large ensembles will decay on similar timescales to those calculated for $N=10$ in the main text.

\begin{figure}[t]
    \centerline{\includegraphics[width=\linewidth]{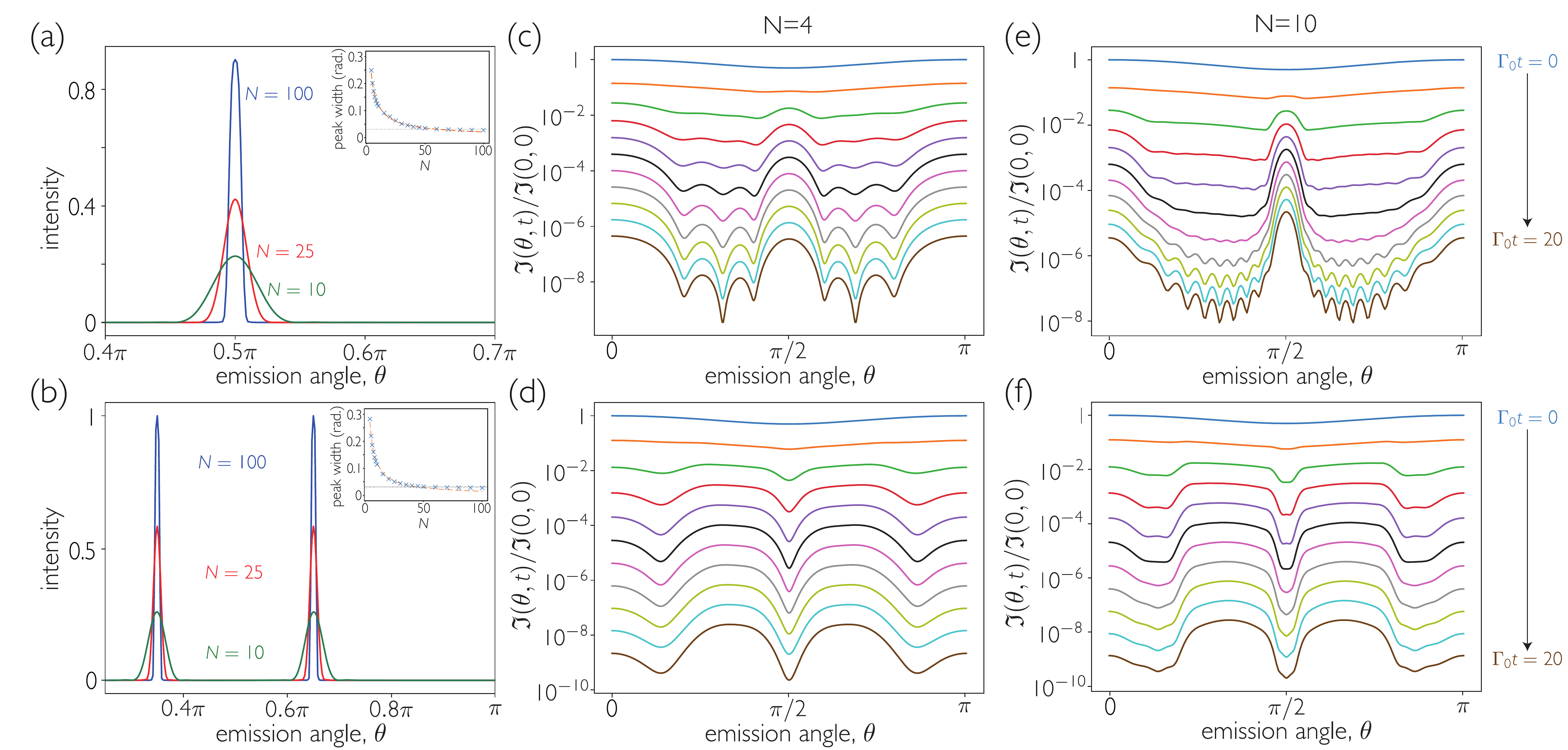}}
\caption{Scaling of directional emission with atom number. Arrays have inter-atomic spacing (top row)~$d/\lambda_0 = 0.9$ and (bottom)~$d/\lambda_0 = 1.1$. \textbf{(a,b)}~Angular emission pattern (in arbitrary units) following the action of the most subradiant operator on a fully inverted array. \textbf{(insets)}~Full width at half maximum of Gaussian fits to emission peaks. Dashed lines are a power law fit $\propto N^{-0.892}$ for $d/\lambda_0=1.1$ and $\propto N^{-0.750}$ for $d/\lambda_0=0.9$. Dotted line is the detector width, $\Delta\theta=0.01\pi$. \textbf{(c-f)} Intensity emitted from a fully-inverted chain of atoms normalized by value at $t=0$. Each curve represents evenly spaced snapshots of the intensity profile for $\Gamma_0t \in [0, 20]$. In all cases, intensity is calculated as measured by detectors of width $\Delta\theta = 0.01\pi$ and atoms are polarized perpendicular to the array.\label{SIFig8}}
\end{figure}

Directional emission due to the action of each jump operator becomes increasingly narrow as the atom number increases. As shown in the insets of Fig.~\ref{SIFig8}(a,b), the width of the emission peaks falls as a power law with $N$ until the detector is too wide to resolve the peak. For small numbers of atoms with inter-atomic separation $d=0.9\lambda_0$, there is a single subradiant operator which is predominantly responsible for late emission. This means that the narrowing of the peak can be observed in the directional emission, as shown in Fig.~\ref{SIFig8}(c,e). For larger atom numbers, more operators become subradiant with increasing $N$ such that the emission profile will not narrow. Figure~\ref{SIFig8}(d,f) show that for $d=1.1\lambda_0$ the emission profile does not substantially change with atom number, even for small atom numbers, as late emission involves multiple similarly subradiant operators.

\color{black}

\subsection{3\;\;\;Non-exponential decay with different initial states}

\subsubsection{3.1\;\;\;Coherent spin states}

Partially excited coherent spin states exhibit non-exponential decay, as discussed in the main text. There, we consider the decay of coherent spin states of the form
\begin{equation}
\ket{\varphi,\mathbf{k}} = \prod\limits_{i=1}^N \sqrt{1-\varphi} \ket{g_i} + \mathrm{e}^{i\mathbf{k}\cdot\rb_i} \sqrt{\varphi}  \ket{e_i}\label{coherenteqnSI}
\end{equation}
where $\mathbf{k}$ is the wavevector of the field that excites the atoms. We can characterize the non-exponential decay by calculating and comparing two fitted decay rates, $\gamma_{\mathrm{early}}$ for an initial period, and $\gamma_{\mathrm{late}}$ for a later period. The periods are chosen such that the population in the late period is small enough to exhibit subradiant decay, yet significant enough for efficient measurement. Here, we fit the decay of the emitted intensity, though similar results are obtained by fitting the excited state population. The initial decay rate is larger than the decay rate at late times in all cases. The initial states have multiple excitations, thus, the action of superradiant operators is enhanced. At late times, completely superradiant paths are fully depleted. Only paths involving subradiant operators remain, with reduced global decay rate.

\begin{figure*}[h!]
\centerline{\includegraphics[width=.85\linewidth]{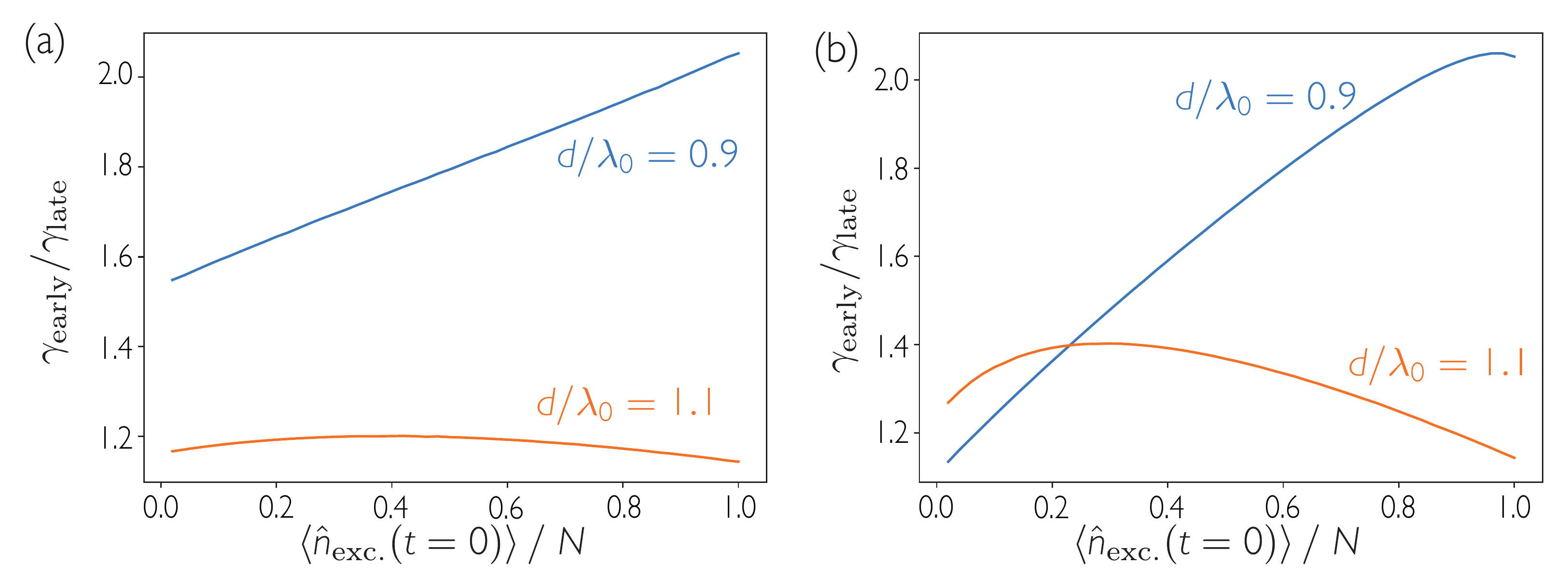}}
\caption{Extracted non-linear decay rates for coherent spin states. Early and late fits, and their ratio, for the decay of the emitted photon pulse from an initial coherent spin state \textbf{(a)}~$\ket{\varphi,k_0\hat{z}}$ and \textbf{(b)}~$\ket{\varphi,k_0\hat{x}}$ as defined by Eq.~\eqref{coherenteqnSI} in the text. In all plots, $N=10$. For $d=0.9\lambda_0$, $\gamma_{\mathrm{early}}$ is fitted for $\Gamma_0t \in [0,4]$ and $\gamma_{\mathrm{late}}$ for $\Gamma_0t \in [10,20]$. For $d=1.1\lambda_0$, $\gamma_{\mathrm{early}}$ is fitted for $\Gamma_0t \in [0,2]$ and $\gamma_{\mathrm{late}}$ for $\Gamma_0t \in [5,10]$. \label{SIFig1}}
\end{figure*}

The early decay rates depend strongly on both the lattice constant and the excitation number, as shown in Fig.~\ref{SIFig1}(a). For $d=0.9\lambda_0$, the early rate is subradiant for low initial populations and increases as $\varphi \rightarrow 1$. For $d=1.1\lambda_0$, the early rate is superradiant for low initial populations, and decreases as $\varphi$ is increased. When all the atoms are fully inverted, there is no coherence in the system and the decay rate cannot depend on the distance. In both cases, the late fit is subradiant without strong dependence on $\varphi$. For $d=0.9\lambda_0$, the ratio between the two fits is initially low and increases as $\varphi$ is increased. For $d=1.1\lambda_0$, the contrast instead peaks at low $\varphi$ and decreases as $\varphi \rightarrow 1$. The ratio is dictated by the set of decay rates for the system. Above and below the geometric resonances, the set of decay rates has a very different profile. This leads to the increased ratio for $d=0.9\lambda$, as the most subradiant decay rates are significantly smaller at that distance than at $d=1.1\lambda_0$. The same idea carries over to all resonances at $d=n\lambda_0/2$.

The states considered in Fig.~\ref{SIFig1}(a,b) are created by a drive propagating along the axis of the array. However, non-exponential decay can still be observed in different drive orientations. Figure~\ref{SIFig1}(c) shows the ratio of measured rates for a perpendicular drive. In particular, the non-exponential decay with $d=0.9\lambda_0$ still has strong contrast.

\subsubsection{3.2\;\;\;Collectively driven states}

\begin{figure}[t!]
\centerline{\includegraphics[width=0.5\linewidth]{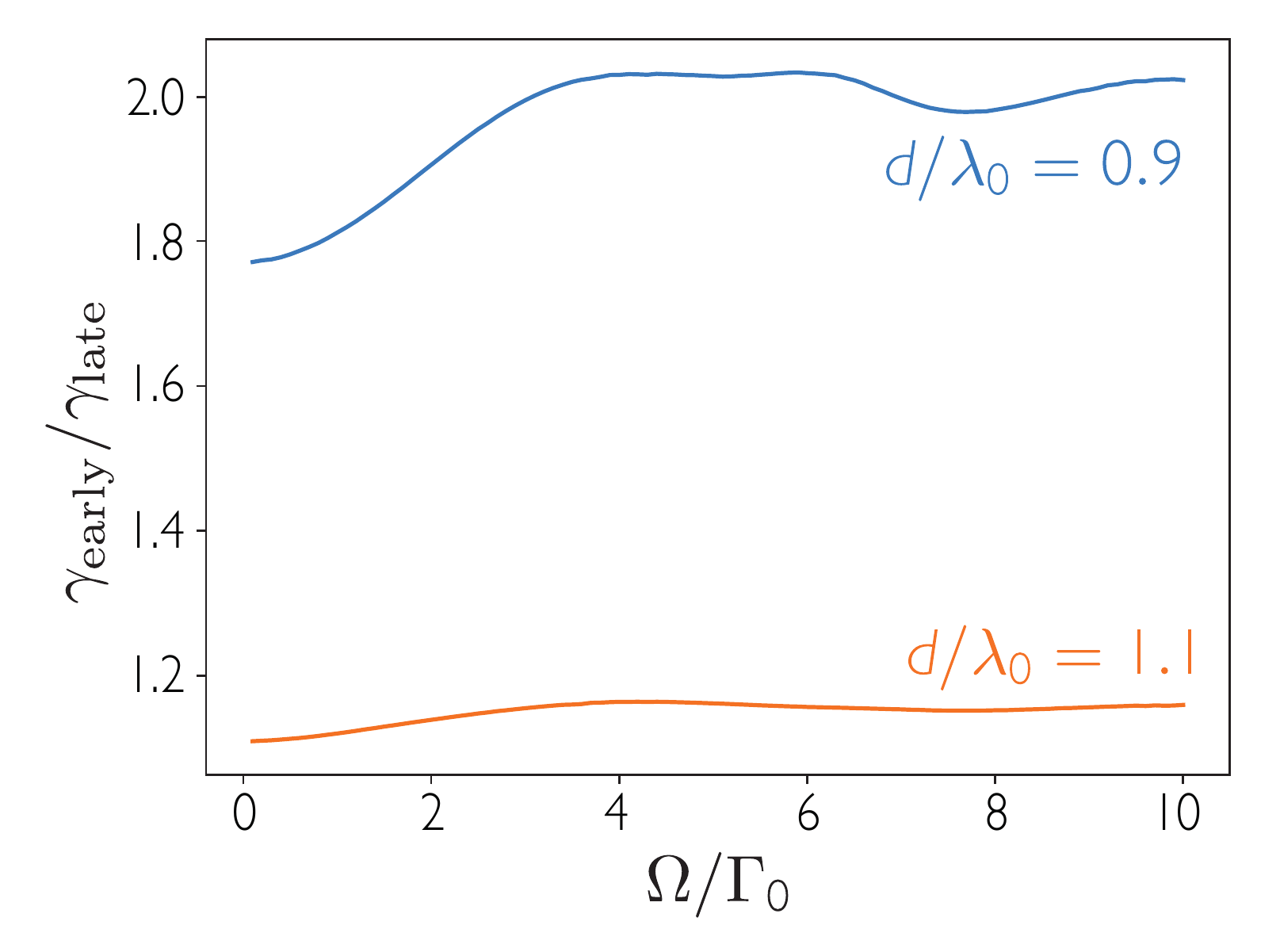}}
\caption{Extracted non-linear decay rates for collectively driven states of $N=8$ atoms. Ratio of early and late fits for the decay of the photon emission pulse induced by a $\sigma^+$-polarized field propagating along the axis of the chain with Rabi frequency $\Omega$ applied for $\Gamma_0\tau_D=1$. For $d=0.9\lambda_0$, $\gamma_{\mathrm{early}}$ is fitted for $\Gamma_0t \in [0,4]$ and $\gamma_{\mathrm{late}}$ for $\Gamma_0t \in [10,20]$. For $d=1.1\lambda_0$, $\gamma_{\mathrm{early}}$ is fitted for $\Gamma_0t \in [0,2]$ and $\gamma_{\mathrm{late}}$ for $\Gamma_0t \in [5,10]$. \label{SIFig3}}
\end{figure}

A longer, weaker drive excites collective modes of the array. We consider the array subject to a time-dependent drive of Rabi frequency $\Omega(t)$, frequency $\omega = \omega_0$ resonant with one of the cycling transitions, and wave-vector $\mathbf{k} = k_0\hat{z}$ where $k_0 = \omega_0 / c = 2\pi / \lambda_0$, with $\lambda_0$ being the atomic transition wavelength. The Hamiltonian in the rotating frame of the drive is given by
\begin{equation}\label{hamdrive}
\mathcal{H}= \hbar\sum_{i,j=1}^N J^{ij}\hat{\sigma}_{eg}^i\hat{\sigma}_{ge}^j + \frac{\hbar\Omega(t)}{2} \sum\limits_{i = 1}^N \left( \mathrm{e}^{ikz_i} \hat{\sigma}_{eg}^i + \mathrm{e}^{-ikz_i}\hat{\sigma}_{ge}^i \right).
\end{equation}
We consider $\Omega(t)$ to be a step function, i.e., on or off. The drive is on for a period $\tau_D$, introducing excited-state population and coherence between ground and excited states into the array. $\tau_D$ is chosen such that significant population is induced, but not so long that the system reaches a steady state. At $t=0$, the drive is instantaneously turned off and the population decays.

The non-exponential nature of the collective decay is also exhibited by collectively driven arrays. In Fig.~\ref{SIFig3}, we plot the ratio of early and late fits as we increase the Rabi frequency of the drive. The ratio is generally more pronounced for larger $\Omega$. However, the dependence on drive strength is non-trivial, as the drive induces oscillations between ground and excited states, and $\braket{\nexc(t=0)}$ does not monotonically increase with $\Omega$. The ratio is again much larger for lattice constants $d = n\lambda_0/2 - \epsilon$.

\subsubsection{3.3\;\;\;``Memory'' of the initial state}

\begin{figure}[h!]
\centerline{\includegraphics[width=.8\linewidth]{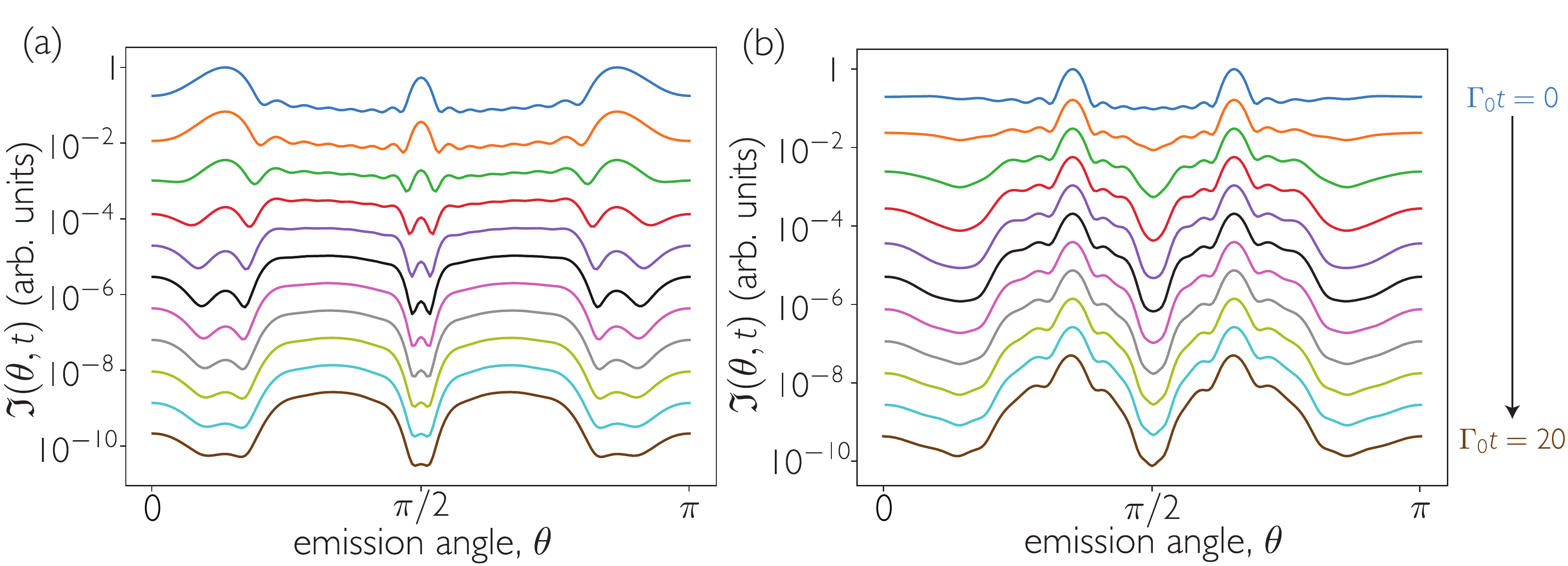}}
\caption{Directional decay of coherent spin states at $d=1.1\lambda_0$. Intensity, normalized by peak intensity at $t=0$, measured by detectors in far-field of angular width $\Delta\theta = 0.01\pi$, from arrays of $N=8$ atoms prepared in a coherent spin state with $\varphi=0.5$ with \textbf{(a)} $\mathbf{k}\cdot(\rb_i-\rb_{i+1}) = 0$ and \textbf{(b)} $\mathbf{k}\cdot(\rb_i-\rb_{i+1}) = 0.5\pi$.\label{SIFig2}}
\end{figure}

A drive that partially excites the array imprints phase relations, and thus correlations, between atoms. These correlations enhance the rates of particular operators, which can be seen in the directional emission, as shown in Fig.~\ref{SIFig2}. By changing the angle of the drive field, different phase patterns can be produced, enhancing the action of different operators. This allows for a different path of evolution, with the likelihoods of each de-excitation path now determined by how the state was prepared as well as the geometry of the array. Some correlations induced by the drive persist throughout the evolution, whereas some are ``washed out''. In the latter case, shown in Fig.~\ref{SIFig2}(a), superradiant decay channels are enhanced by the drive. Such decay occurs fast, and does not impact evolution at late times. Conversely, in Fig.~\ref{SIFig2}(b), the drive enhances subradiant decay, and the enhanced peak remains superimposed over the subradiant background at late times.

\subsection{4\;\;\;Role of imperfections}

Our results are robust under reasonable experimental imperfections. In this section, we study the impact of finite filling fraction and disorder in the atomic positions (i.e., classical disorder).

Creating completely defect-free atomic arrays is a complex task that has seen remarkable progress in recent years. Atom arrays are generally loaded using light-assisted collisions, where atoms are lost in pairs, such that each site is loaded with zero or one atom with approximately equal probability~\cite{Schlosser02}. Careful tuning of the collisions allows for more efficient filling~\cite{Grunzweig10,Fung15,Lester15,Brown19}, while imaging and rearranging an array has been shown to bring the filling fraction close to unity~\cite{Kim16,Barredo16,Endres16,Barredo18,Kumar18}. Nevertheless, these techniques are unnecessary to observe the phenomena described in this manuscript.

Position disorder also generates imperfection in the array. Here, we consider three-dimensional Gaussian positional noise with standard deviations [$\mathbf{\Xi}_\mathrm{lattice}/\lambda_0 = \set{0.1,0.1,0.04}$] in the $\{x,y,z\}$-axis, slightly larger than those quoted in a recent optical lattice experiment~\cite{Rui20}. Such disorder technically breaks the assumption of atoms behaving as two-level systems, as the re-scattered field at the atomic positions has multiple polarization components when the atoms are not perfectly aligned. These components drive other transitions, and it is necessary to consider the full hyperfine structure of the atoms~\cite{Asenjo19}. It is possible to return to a two-level approximation by applying a strong magnetic field, such that all but the relevant cycling transition are detuned due to Zeeman shifts.

\begin{figure}[h!]
    \centerline{\includegraphics[width=.95\linewidth]{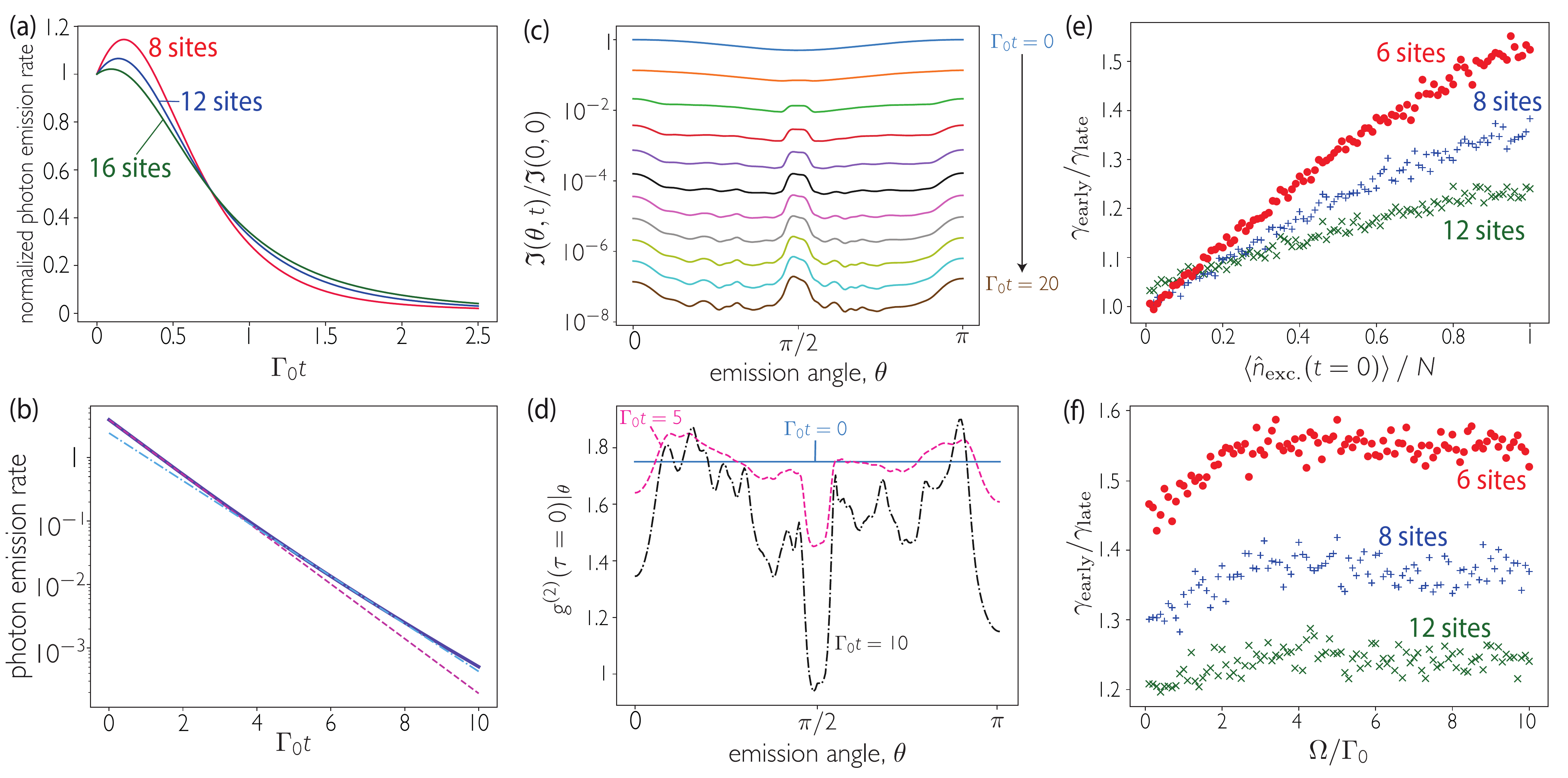}}
\caption{Role of imperfections on the collective decay of a 1D array. In all cases, presented data are averages over 50 randomly generated arrays. \textbf{(a)}~Rate of photon emission normalized by atom number [$-\mathrm{d}\bnexc/\mathrm{d}(N\Gamma_0t)$] of an array of $N=8$ atoms arranged in arrays with inter-site distance $d=0.1\lambda_0$ and different numbers of sites. \textbf{(b)}~Decay of initial coherent spin states $\ket{\varphi=0.5,k_0\hat{z}}$ of $N=8$ atoms arranged in a lattice of 16 sites and three-dimensional noise $\mathbf{\Xi}_{\mathrm{lattice}}$ applied. Comparison is made to exponential fits of the dynamics over (dashed) $\Gamma_0t \in [0,2]$ and (dot-dashed) $\Gamma_0t \in [5,10]$. \textbf{(c,d)}~ Directional emission from a fully-inverted array of $N=8$ atoms arranged in 16 sites with three-dimensional noise $\mathbf{\Xi}_\mathrm{lattice}$ applied. (c)~Intensity, normalized by peak intensity at $t=0$, measured by detectors in far-field of angular width $\Delta\theta=0.01\pi$. Each curve represents evenly spaced snapshots of the intensity profile between $\Gamma_0t = 0\rightarrow20$. (d)~Directional second order correlation function. \textbf{(e,f)}~Ratio of average early and late fitted decay rates for the photon emission from different size arrays randomly filled with $N=6$ atoms with three-dimensional noise $\mathbf{\Xi}_\mathrm{lattice}$ applied. $\gamma_{\mathrm{early}}$ is fitted for $\Gamma_0t \in [0,4]$ and $\gamma_{\mathrm{late}}$ for $\Gamma_0t \in [10,20]$. In (e), initial states are coherent spin states as defined by Eq.~\eqref{coherenteqnSI} in the text. In (f), the array is driven for $\Gamma_0\tau_D = 1$ by a field of Rabi frequency $\Omega$ propagating along $\hat{z}$ with $\sigma^+$-polarization. In (b-f), the array has inter-site distance $d=0.9\lambda_0$. In all cases, atoms are polarized perpendicular to the array.
\label{SIFig5}}
\end{figure}

Results from the main text can be reproduced in the presence of these imperfections, as shown in Fig.~\ref{SIFig5}. The superradiant burst remains for an array of $d=0.1\lambda_0$ even when the array is half-filled, as shown in Fig.~\ref{SIFig5}(a). Experimentally accessible arrays, with partial filling fraction and positional noise, still exhibit non-exponential decay with significant directional properties. Fig.~\ref{SIFig5}(b-f) show that while the contrast is reduced by the imperfections, all of the features still remain.

\end{document}